\documentclass[12pt]{article}
\pdfoutput=1

\usepackage{jheppub}
\usepackage{amssymb,amsfonts,amsmath}
\usepackage{enumerate}
\usepackage{mathrsfs}
\usepackage{tikz}
\usepackage[section]{placeins}
\usepackage{amsthm}

\allowdisplaybreaks

\newcommand{\be}{\begin{equation}}
\newcommand{\ee}{\end{equation}}
\newcommand{\bea}{\begin{eqnarray}}
\newcommand{\eea}{\end{eqnarray}}

\newcommand{\CA}{\mathcal{A}}
\newcommand{\CB}{\mathcal{B}}
\newcommand{\CC}{\mathcal{C}}
\newcommand{\CD}{\mathcal{D}}
\newcommand{\CF}{\mathcal{F}}

\newcommand{\CH}{\mathcal{H}}

\newcommand{\CO}{\mathcal{O}}

\newcommand{\CR}{\mathcal{R}}
\newcommand{\CT}{\mathcal{T}}
\newcommand{\CU}{\mathcal{U}}
\newcommand{\CV}{\mathcal{V}}

\newcommand{\CZ}{\mathcal{Z}}

\newcommand{\lr}{\left (}
\newcommand{\rr}{\right )}
\newcommand{\ls}{\left [}
\newcommand{\rs}{\right ]}

\newcommand\qt\tau

\newcommand{\p}{\partial}

\renewcommand{\tilde}[1]{\widetilde{#1}}

\makeatletter
\renewcommand{\@seccntformat}[1]{\csname the#1\endcsname.\,\,}
\makeatother
\allowdisplaybreaks

\newtheorem{Def}{Identity}

\let \savenumberline \numberline
\def \numberline#1{\savenumberline{#1.}}

\makeatletter
\def\@fpheader{\relax}
\makeatother

\def\bea{\begin{eqnarray}}
\def\eea{\end{eqnarray}}

\title{\ \vspace{1.6cm} \\
\scalebox{0.8}{Background Field Method for Nonlinear Sigma Models}
\scalebox{0.8}{in Nonrelativistic String Theory}}
\author{Ziqi Yan and Matthew Yu}
\emailAdd{zyan@perimeterinstitute.ca}
\emailAdd{myu@perimeterinstitute.ca}
\affiliation{
        Perimeter Institute for Theoretical Physics\\
        31 Caroline St N, Waterloo, ON N2L 2Y5, Canada}
\abstract{We continue the study of nonrelativistic string theory in background fields. Nonrelativistic string theory is described by a nonlinear sigma model that maps a relativistic worldsheet to a non-Lorentzian and non-Riemannian target space geometry, which is known to be string Newton-Cartan geometry. We develop the covariant background field method in this non-Riemannian geometry. We apply this background field method to compute the beta-functions of the nonlinear sigma model that describes nonrelativistic string theory on a string Newton-Cartan geometry background, in presence of a Kalb-Ramond two-form and dilaton field.}

\begin{document}

\maketitle

\section{Introduction}

Nonrelativistic string theory \cite{Gomis:2000bd, Danielsson:2000gi, Klebanov:2000pp} is a unitary and ultraviolet finite quantum theory of nonrelativistic gravity. This theory
has a spectrum of string excitations with a (string)-Galilean invariant dispersion relation and a well-behaved perturbative expansion.
Nonrelativistic string theory in a curved background is described by a two-dimensional relativistic nonlinear sigma model (NLSM) on the worldsheet, which contains additional worldsheet fields beyond those that describe the motion of strings in spacetime. Such additional fields play the role of Lagrange multipliers in the NLSM, and
endow a two-dimensional foliation structure in spacetime.  This foliated spacetime has a longitudinal and transverse sector related to each other by the (string-)Galilean boost,
which is nontrivially realized in the worldsheet NLSM action. The appropriate string-Galilean boost invariant spacetime geometry that nonrelativistic string theory couples to is the string Newton-Cartan geometry \cite{stringyNC, Bergshoeff:2018yvt, stringyNClimit}.\,\footnote{For other recent work on nonrelativistic strings, see \cite{Batlle:2016iel, Gomis:2016zur, Batlle:2017cfa, HHO, Kluson, Kluson:2018grx, Harmark:2018cdl, Kluson:2018vfd, Kluson:2019ifd, Roychowdhury:2019qmp, Roychowdhury:2019vzh, Gallegos:2019icg, Harmark:2019upf, Roychowdhury:2019olt, Roychowdhury:2019sfo}. In \cite{HHO, Harmark:2018cdl, Harmark:2019upf}, with the zero torsion condition $d\tau = 0$, a specific truncation of the string Newton-Cartan gravity (with zero $B$-field and dilaton) in the target space was considered, which leads to Newton-Cartan gravity in one dimension lower, supplemented with an extra worldsheet scalar parametrizing the spatial foliation direction. A more thorough examination of this relation in the presence of the $B$-field was later explored in \cite{Harmark:2019upf}, where the necessity of imposing the torsionless constraint $D_{[\mu} \tau_{\nu]}{}^A = 0$ in nonrelativistic string theory was studied at the classical level. We will focus on the zero torsion case in the bulk of the paper, and only comment on the quantum effects from a nonzero torsion in Appendix \ref{app:torsion}.}\,\footnote{See \cite{Pereniguez:2019eoq} for recent results on $p$-brane Newton-Cartan geometry.} This target space geometry does not admit any Riemannian or Lorentzian metric, and therefore is intrinsically different from Riemannian geometry. 

The two-dimensional quantum NLSM that describes nonrelativistic string theory in a curved background contains spacetime background fields such as the string Newton-Cartan gauge fields, $B$-field and dilaton. These spacetime fields play the role of classically marginal couplings. The absence of any worldsheet Weyl anomaly at the quantum level in nonrelativistic string theory requires tuning these background fields such that 
all beta-functions of the worldsheet NLSM vanish, leading to a two-dimensional conformal field theory. The vanishing  beta-functions give rise to equations of motion that determine the backgrounds on which nonrelativistic string theory can be consistently defined quantum mechanically.

The spacetime equations of motion in nonrelativistic string theory have recently been derived in \cite{Gomis:2019zyu} by first computing the linearized equations that follow from the Weyl invariance condition of the vertex operators around flat spacetime, and then going to higher orders in the background perturbations.\,\footnote{For a study of Weyl invariance of the sigma model in a torsional Newton-Cartan background, see \cite{Gallegos:2019icg}.} The equations of motion found in \cite{Gomis:2019zyu} are to nonrelativistic string theory what the supergravity equations of motion are to relativistic string theory.
The method used in \cite{Gomis:2019zyu} has the benefit of being straightforward, and the origin of the spacetime equations of motion is physically manifest when directly dealing with the vertex operators. The result in \cite{Gomis:2019zyu} is then corroborated in a companion paper \cite{stringyNClimit} by taking a subtle limit of beta-functions in relativistic string theory.

Since the vertex operators are defined around flat spacetime, the method in \cite{Gomis:2019zyu} is not covariant with respect to the NLSM reparametrizations. In principle, it is straightforward to compute all higher order terms in the equations of motion by evaluating various operator product expansions, but in practice, it is more feasible to determine these higher order terms by requiring that all ingredients are covariantized. It is therefore of immediate interest to develop a covariant background field method in string Newton-Cartan geometry and apply it to derive the beta-functions in the NLSM action that describes nonrelativistic string theory, in which we compute the Weyl anomalies by using Feynman diagrams.\,\footnote{For the same calculation in relativistic string theory, see \cite{Callan:1985ia, Howe:1986vm, Callan:1989nz, Ketov:2000dy} for some useful references.} This not only provides us with a powerful way for extracting higher order contributions in background fields, complementary to the method used in \cite{Gomis:2019zyu}, but also improves our understanding of string Newton-Cartan geometry.
This could facilitate future studies when more structures such as supersymmetry and D-branes are included.

In this paper, we develop the covariant background field method for string Newton-Cartan geometry on a non-Riemannian manifold.\,\footnote{The background field method for the sigma model in torsional Newton-Cartan geometry has been discussed in \cite{Gallegos:2019icg}, which has a different starting point from what we do. See more in Appendix \ref{app:torsion}.} In this geometry, an affine connection can be defined in the absence of any metric. As a result, a notion of geodesic can be introduced in string Newton-Cartan geometry, by which we can define a nonlinear background quantum splitting for the nonrelativistic string theory NLSM. This procedure allows us to achieve covariance through the calculation of beta-functions. 

The paper is organized as follows. In \S\ref{sec:review}, we first review basic ingredients of the NLSM that describes nonrelativistic string theory, and then develop the covariant background field method on string Newton-Cartan geometry in \S\ref{sec:cobackground}, where we also give the covariant expansion of the NLSM. In \S\ref{sec:quancal}, we study the one-loop effective action for the NLSM without the dilaton and derive the beta-functions at the leading $\alpha'$-order. Ensuring that the NLSM is well-defined when taking into account higher loop corrections requires a nonrenormalization theorem that we prove in \S\ref{sec:nonre}. In \S\ref{sec:dilaton}, we consider the contributions from the dilaton. Finally, we give the conclusion in \S\ref{sec:conclusions}. In Appendix \ref{app:ids}, we present identities in string Newton-Cartan geometry that are useful for the derivation of beta-functions. In Appendix \ref{app:torsion}, we discuss the quantum effects in the NLSM from turning on torsion in string Newton-Cartan geometry. 

\section{The Classical Action} \label{sec:classicalaction}

\subsection{Sigma model on string Newton-Cartan background} \label{sec:review}

Let $\mathcal{M}$ be a $d$ dimensional manifold, parametrized by coordinates $x^\mu$, $\mu = 0, 1, \cdots, d-1$\,. In the following, we give a brief review of string Newton-Cartan geometry on this manifold $\mathcal{M}$ \cite{stringyNC, stringyNClimit}. Let $\mathcal{T}_p$ be the tangent space attached to a point $p$ in $\mathcal{M}$\,. Decompose $\mathcal{T}_p$ into two longitudinal directions with coordinates $x^A$\,, $A = 0, 1$ and $d-2$ transverse directions with coordinates $x^{A'}$\,, $A' = 2, \cdots, d-1$\,. Endow $\mathcal{M}$ with a two-dimensional foliation structure by introducing a longitudinal Vielbein field $\tau_\mu{}^A$ and a transverse Vielbein field $E_\mu{}^{A'}$\,. The inverse longitudinal (transverse) Vielbein fields $\tau^\mu{}_A$ ($E^\mu{}_{A'}$) satisfy the following invertibility conditions:
\begin{subequations} \label{eq:invertibility}
\begin{align}
	\tau^\mu{}_A \, \tau_\mu{}^B & = \delta_A^B\,, 
		&
	\tau^\mu{}_{\!A} \, \tau_\nu{}^A + E^\mu{}_{A'} E_\nu{}^{A'} & = \delta^\mu_\nu\,, \\[2pt] 
	E^\mu{}_{A'} E_\mu{}^{B'} & = \delta_{A'}^{B'}\,, 
		&
	\tau^\mu{}_{\!A} \, E_\mu{}^{A'} = E^\mu{}_{\!A'} \, \tau_\mu{}^A & = 0\,.
\end{align}
\end{subequations}
The longitudinal and transverse directions are related by the so-called string-Galilean boost, which acts on the Vielbein fields and their inverses as
\begin{subequations} \label{eq:sGboost}
\begin{align}
	\delta_G \hspace{0.1mm} \tau_\mu{}^A & = 0\,, 
		&
	\delta_G E_\mu{}^{A'} & = - \tau_\mu{}^A \Lambda_A{}^{A'}\,, \\
	\delta_G \hspace{0.1mm} \tau^\mu{}_A & = E^\mu{}_{\!A'} \, \Lambda_A{}^{A'}\,, 
		&
	\delta_G E^\mu{}_{A'} & = 0\,.
\end{align}
\end{subequations}
To construct string Newton-Cartan geometry, one also needs to introduce an extra gauge field $m_\mu{}^A$ that transforms nontrivially under the string-Galilean boost, with
\be \label{eq:boostmG}
	\delta_G \hspace{0.1mm} m_\mu{}^A = E_\mu{}^{A'} \Lambda^A{}_{A'}\,.
\ee
From the gauge fields $\tau_\mu{}^A\,, E_\mu{}^{A'}$ and $m_\mu{}^A$\,, one can construct a boost invariant two-tensor with lowered curved indices (in addition to $\tau_{\mu\nu} \equiv \eta_{AB} \tau_\mu{}^A \tau_\nu{}^B$\,, which is trivially boost invariant),
\be \label{eq:sGboostinv}
	H_{\mu\nu} \equiv E_\mu{}^{A'} E_\nu{}^{A'} + \bigl( \tau_\mu{}^A m_\nu{}^B + \tau_\nu{}^A m_{\mu}{}^B \bigr) \, \eta_{AB}\,.
\ee
We emphasize that $m_\mu{}^A$ is essential for $H_{\mu\nu}$ to be string-Galilean boost invariant.  

In string Newton-Cartan geometry, the gauge field $m_\mu{}^A$ is associated with a noncentral extension generator $Z_A$ in the string Newton-Cartan algebra that underlies string Newton-Cartan geometry \cite{stringyNC, stringyNClimit}. This generator $Z_A$ shows up in the Lie bracket that involves the transverse translation generator $P_{A'}$ and the string-Galilean boost generator $G_{AA'}$ \cite{stringyNC},
\be
	[G_{AA'}\,, P_{B'}] = \delta_{A'B'} Z_A\,.
\ee
The $Z_A$ transformation acts nontrivially on $m_\mu{}^A$ as
\be \label{eq:deltaZ}
	\delta_Z \, m_\mu{}^A = D_\mu \sigma^A\,,
\ee
and it acts trivially on $\tau_\mu{}^A$\,, $E_\mu{}^{A'}$ and $B_{\mu\nu}$\,.\footnote{In \cite{Harmark:2019upf}, a different spacetime gauge symmetry group is proposed, in which there is a different $Z_A$ symmetry transformation that acts nontrivially on both $m_\mu{}^A$ and $B_{\mu\nu}$\,.} Here, $D_\mu$ is covariant with respect to the longitudinal Lorentz boost transformation that acts on the index $A$\,. Note that $H_{\mu\nu}$ defined in \eqref{eq:sGboostinv} is not invariant under the $Z_A$ symmetry, and thus $H_{\mu\nu}$ does not constitute a metric. It is in this sense that string Newton-Cartan geometry is non-Riemannian.  

Next, we define the affine connection $\Gamma$ on $\mathcal{M}$ by imposing the Vielbein postulates
\begin{align} \label{eq:vielbeinpostulates}
    D_\mu \tau_\nu{}^A - \Gamma^\rho{}_{\mu\nu} \, \tau_\rho{}^A = 0\,,
        \quad%
    D_\mu E_\nu{}^{A'} - \Gamma^\rho{}_{\mu\nu} \, E_\rho{}^{A'} = 0\,,
\end{align}
where 
\begin{align} \label{eq:covariantdertauE}
    D_\mu \tau_\nu{}^A & \equiv \p_\mu \tau_\nu{}^A - \epsilon^A{}_B \, \Omega_{\mu} \, \tau_{\nu}{}^B\,, 
        &
    D_\mu E_\nu{}^{A'} & \equiv \p_\mu E_\nu{}^{A'} + \Omega_{\mu}{}^{AA'} \tau_{\nu A} - \Omega_{\mu}{}^{A'B'} E_{\nu}{}^{B'}\,.
\end{align}
We have introduced the spin connections $\Omega_\mu$\,, $\Omega_\mu{}^{AA'}$ and $\Omega_\mu{}^{A'B'}$, which are associated with the longitudinal Lorentz rotation, the string-Galilean boost and the transverse rotation, respectively. 
In string Newton-Cartan geometry, $\Gamma^\rho{}_{\mu\nu}$ is required to satisfy the torsionless condition,
\be \label{eq:torsionlessGamma}
    \Gamma^\rho{}_{\mu\nu} = \Gamma^\rho{}_{\nu\mu}\,.
\ee
This affine connection can be written in terms of various spin connections by using the Vielbein postulates in \eqref{eq:vielbeinpostulates} and the invertibility conditions in \eqref{eq:invertibility} as
\be \label{eq:affineconnectioncoeff}
    \Gamma^\rho{}_{\mu\nu} = \tau^\rho{}^{}_{\!A} \, D^{}_{(\mu} \tau^{}_{\nu)}{}^A + E^\rho{}_{A'} \bigl( \p_{(\mu} E_{\nu)}{}^{A'} + \Omega_{(\mu}{}^{AA'} \tau_{\nu) A} - \Omega_{(\mu}{}^{A'B'} E_{\nu)}{}^{B'} \bigr)\,.
\ee
Note that $\Omega_\mu{}^{AA'}$ and $\Omega_\mu{}^{A'B'}$ contain $m_\mu{}^A$ dependence \cite{stringyNC, stringyNClimit}.
Furthermore, from \eqref{eq:vielbeinpostulates} and \eqref{eq:torsionlessGamma}, we find the curvature constraints
\be \label{eq:curvatureconstraint}
    D^{}_{[\mu} \tau^{}_{\nu]}{}^A = 0\,,
        \qquad
    D^{}_{[\mu} E^{}_{\nu]}{}^{A'} = 0\,.
\ee 
The condition $D_{[\mu} \tau_{\nu]}{}^A = 0$ imposes nontrivial constraints on $\tau_\mu{}^A$\,,
\be
    \epsilon_C{}^{(A} \tau_{[\mu}{}^{B)} \p^{}_\nu \tau_{\rho]}{}^C = 0\,.
\ee
This defines the generalized hypersurface orthogonality condition in a two-dimensional foliation structure.\,\footnote{This is an analogue of the hypersurface orthogonality condition in spacetime with a one-dimensional foliation structure.} This concludes our review on string Newton-Cartan geometry.

We now proceed to the construction of the NLSM that describes nonrelativistic string theory on a general curved string Newton-Cartan background, in presence of a Kalb-Ramond two-form and dilaton field. 
Let $\Sigma$ be a two-dimensional Euclidean space, parametrized by $\sigma^\alpha$\,, $\alpha = 1, 2$\,. Let the spacetime coordinates $x^\mu = x^\mu (\sigma)$ be a mapping from the worldsheet $\Sigma$ to the target space $\mathcal{M}$\,. We also introduce two additional worldsheet fields $\lambda(\sigma)$ and $\overline{\lambda} (\sigma)$\,. 
The NLSM is \cite{Bergshoeff:2018yvt, Gomis:2005pg}
\be \label{eq:smaction}
	S = \frac{1}{4\pi\alpha'} \int d^2 \sigma \, \Bigl\{ \p x^\mu \, \overline{\p} x^\nu \bigl( H_{\mu\nu}[x] + B_{\mu\nu}[x] \bigr) + \lambda \, \overline{\p} x^\mu \, \tau_\mu[x] + \overline{\lambda} \, \p x^\mu \, \overline{\tau}_\mu[x] \Bigr\}\,,
\ee
where $B_{\mu\nu}$ is the Kalb-Ramond two-form field and
\begin{subequations}
\begin{align}
	\p & = \frac{\p}{\p \sigma^1} - i \frac{\p}{\p \sigma^2}\,,
		&
	\tau_\mu & = \tau_\mu{}^0 + \tau_\mu{}^1\,, \\[2pt]
	\overline{\p} & = \frac{\p}{\p \sigma_1} + i \frac{\p}{\p \sigma_2}\,, 
		&
	\overline{\tau}_\mu & = \tau_\mu{}^0 - \tau_\mu{}^1\,.
\end{align}
\end{subequations}
Since the worldsheet is taken to be flat, there is no dilaton in the action. This sigma model action is invariant under the string Newton-Cartan gauge symmetries \cite{Bergshoeff:2018yvt}. 
In order for the sigma model action \eqref{eq:smaction} to be invariant under the $Z_A$ gauge symmetry, one also needs to transform the Lagrange multipliers $\lambda$ and $\overline{\lambda}$ as
\be
	\delta_Z \lambda = \p x^\mu \, D_\mu \bigl( \sigma^0 - \sigma^1 \bigr)\,,
		\qquad
	\delta_Z \overline{\lambda} = \bar{\p} x^\mu \, D_\mu \bigl( \sigma^0 + \sigma^1 \bigr)\,;
\ee
in addition, one needs to impose the hypersurface orthogonality condition $D_{[\mu} \tau_{\nu]}{}^A = 0$ in \eqref{eq:curvatureconstraint}. This condition prohibits the $\lambda \overline{\lambda}$ operator from being generated by quantum corrections in the effective action, which would otherwise deform the theory towards relativistic string theory \cite{Gomis:2019zyu}.\,\footnote{See more in \S\ref{sec:nonre} and Appendix \ref{app:torsion}.}

Finally, it will prove later to be useful in the quantum calculations to take the following field redefinition of $\lambda$ and $\overline{\lambda}$ in the sigma model action \eqref{eq:smaction}:
\be \label{eq:lambdalambdatrans}
	\lambda \rightarrow \lambda - \p x^\mu \, \bigl( m_\mu{}^0 - m_\mu{}^1 \bigr)\,,
		\qquad
	\overline{\lambda} \rightarrow \overline{\lambda} - \overline{\p} x^\mu \, \bigl( m_\mu{}^0 + m_\mu{}^1 \bigr)\,. 
\ee
Consequently, \eqref{eq:smaction} can equivalently be written as
\be \label{eq:eqsmaction}
	S = \frac{1}{4\pi\alpha'} \int d^2 \sigma \, \Bigl\{ \p x^\mu \, \overline{\p} x^\nu \bigl( E_{\mu\nu}[x] + A_{\mu\nu}[x] \bigr) + \lambda \, \overline{\p} x^\mu \tau_\mu[x] + \overline{\lambda} \, \p x^\mu \overline{\tau}_\mu[x] \Bigr\}\,,
\ee
where
\be
	E_{\mu\nu} \equiv E_\mu{}^{A'} E_\nu{}^{A'}\,,
		\qquad
	A_{\mu\nu} \equiv B_{\mu\nu} + \bigl( m_\mu{}^A \tau_\nu{}^B - m_\nu{}^A \tau_\mu{}^B \bigr) \epsilon_{AB}\,.
\ee 
Note that $E_{\mu\nu}$ is invariant under the $Z_A$ gauge transformation but not invariant under the string-Galilean boost, and $A_{\mu\nu}=-A_{\nu\mu}$\,.
The associated path integral is
\be \label{eq:pathintegral}
    \mathcal{Z} = \int \mathscr{D} \lambda \, \mathscr{D} \overline{\lambda} \, \mathscr{D} x^\mu \sqrt{G} \, e^{-S}\,,
\ee
Here, $G$ is given by \cite{Gomis:2019zyu, stringyNClimit}
\be \label{eq:det}
    G \equiv \det^{(d)} \bigl( H_{\mu\nu} \bigr) \, \det^{(2)} \! \big( \tau_\rho{}^{A} H^{\rho\sigma} \tau_\sigma{}^{B} \bigr)\,,
\ee
where $H^{\mu\nu}$ is the inverse of $H_{\mu\nu}$ such that $H^{\mu\rho} H_{\rho\nu} = \delta^\mu_\nu$\,. Note that $G$ is independent of $m_\mu{}^A$\,. 
The action \eqref{eq:eqsmaction} and its path integral \eqref{eq:pathintegral} will be what we mostly work with in the rest of the paper. 
 
\subsection{Covariant background field method} \label{sec:cobackground}

In this section, we develop the background field method in string Newton-Cartan geometry and apply it to expand the sigma model defined in \eqref{eq:eqsmaction} around a covariant background.

Consider a sufficiently small neighborhood $\mathcal{O}$ of a point $x_0^\mu$ in $\mathcal{M}$\,. For an arbitrary point $x^\mu$ in $\mathcal{O}$\,, there exists a unique geodesic interpolating between $x_0^\mu$ and $x^\mu$\,, parametrized by $y^\mu(s)$\,, with an affine parameter $s \in [0,1]$ , such that
\be \label{eq:geodesicy}
	\frac{d^2 y^\mu (s)}{d s^2} + \Gamma^\mu{}_{\rho\sigma} [y(s)] \, \frac{d y^\rho (s)}{ds} \frac{d y^\sigma (s)}{ds} = 0\,,
\ee
and
\be
	y^\mu(0) = x_0^\mu\,,
		\qquad
	y^\mu(1) = x^\mu\,.
\ee
Define
\be \label{eq:ellmudef}
	\ell^\mu \equiv \frac{d y^\mu(s)}{ds} \bigg|_{s=0}\,, 
\ee
which is the tangent vector to the geodesic at $s=0$\,. Here, the connection coefficients $\Gamma^\rho{}_{\mu\nu}$ are given in \eqref{eq:affineconnectioncoeff}. Since we have relocated the dependence on $m_\mu{}^A$ to be in the two-form field $A_{\mu\nu}$ in \eqref{eq:eqsmaction}, the string Newton-Cartan geometry now effectively has $m_\mu{}^A=0$\,. This implies that we also need to set $m_\mu{}^A = 0$ in $\Gamma$ when later applying this background field method to expand the action \eqref{eq:eqsmaction}.

Next, we define a covariant derivative $\nabla_{\!s}$ for the affine parameter $s$\,, which acts respectively on covariant vector $V_\mu$ and contravariant vector $V^\mu$ as
\be
	\nabla_{\!s} \, V_\mu = \frac{dV_\mu}{ds} - \Gamma^\rho{}_{\mu\sigma} \, V_\rho \, \frac{dy^\sigma(s)}{ds} \,, 
		\qquad
	\nabla_{\!s} V^\mu = \frac{dV^\mu}{ds} + \Gamma^\mu{}_{\rho\sigma} \, V^\rho \, \frac{dy^\sigma(s)}{ds} \,.
\ee
Here, $\nabla_{\!\mu}$ is the covariant derivative defined with respect to the affine connection $\Gamma$\,. 
%
%
Note that the differential $d/ds$ acting on $\mathcal{M}$ commutes with the worldsheet derivative $\p_\alpha \equiv \p / \p \sigma^\alpha$\,. Therefore,
\begin{subequations} \label{eq:idforexpansions}
\begin{align}
    \nabla_{\!s} \, \p_\alpha y^\mu(s) 
    & = \nabla_{\!\alpha} \frac{dy^\mu(s)}{ds}\,,
        \qquad%
    \nabla_{\!s} \frac{d y^\mu(s)}{d s} = 0\,, \\[2pt]
    \bigl[ \nabla_s\,, \nabla_\alpha \bigr] \frac{dy^\mu (s)}{ds} & = \p_\alpha y^\nu(s) \, R^\mu{}_{\rho\sigma\nu} [y(s)] \, \frac{dy^\rho (s)}{ds} \, \frac{dy^\sigma (s)}{ds}\,,
\end{align}
\end{subequations}
where
\be \label{eq:RGamma}
    R^\mu{}_{\rho\sigma\nu} = \p_\sigma \Gamma^\mu{}_{\nu\rho} - \p_\nu \Gamma^\mu{}_{\sigma\rho} + \Gamma^\mu{}_{\sigma\kappa} \Gamma^\kappa{}_{\nu\rho} - \Gamma^\mu{}_{\nu\kappa} \Gamma^\kappa{}_{\sigma\rho}
\ee
defines the curvature tensor. 

We expand the Lagrangian associated with the sigma model action in \eqref{eq:eqsmaction} around the classical background field $x^\mu_0 (\sigma)$\,, with respect to the affine parameter $s$. We take the fluctuation field to be the tangent vector $\ell^\mu (\sigma)$ defined in \eqref{eq:ellmudef}. Furthermore, we define
\begin{align} \label{eq:Sys}
    S [y(s)] = \frac{1}{4\pi\alpha'} \int \! d^2 \sigma \, \Bigl\{ \p y^\mu(s) \, \overline{\p} y^\nu(s) & \, \bigl( E_{\mu\nu}[y(s)] + A_{\mu\nu}[y(s)] \bigr) \notag \\
        & + \lambda \, \overline{\p} y^\mu (s) \, \tau_\mu[y(s)] + \overline{\lambda} \, \p y^\mu (s) \, \overline{\tau}_\mu[y(s)] \Bigr\}\,,
\end{align}
such that $S[y(1)] = S[x]$ in \eqref{eq:eqsmaction}. Then, 
\be \label{eq:S[x]}
    S [x] = S [x_0] + S^{(1)} [x_0\,, \ell] + S^{(2)} [x_0\,, \ell] + O (\ell^3)\,,
\ee
where
\be
    S^{(n)} [x_0] = \frac{1}{n!} \frac{d^n \! S[y(s)]}{ds^n} \bigg|_{s=0}\,.
\ee
We assumed that the Lagrange multipliers $\lambda$ and $\overline{\lambda}$ are independent of the affine parameter $s$\,.
It then follows that
\begin{align}
    S^{(1)} [x_0] = \frac{1}{4\pi\alpha'} \! \int \! d^2 \sigma \Bigl\{ \nabla \ell^\rho \, \overline{\Gamma}_\rho + \overline{\nabla} \ell^\rho \, \Gamma_\rho + \ell^\rho \, \Delta_\rho & + \lambda \, \bigl( \overline{\nabla} \ell^\rho \, \tau_\rho + \overline{\p} x_0^\mu \, \ell^\rho \, \nabla_{\!\rho} \tau_\mu \bigr) \notag \\
    & + \overline{\lambda} \, \bigl( {\nabla} \ell^\rho \, \overline{\tau}_\rho + {\p} x_0^\mu \, \ell^\rho \, \nabla_{\!\rho} \overline{\tau}_\mu \bigr) \Bigr\}\,,
\end{align}
and
\begin{align}
    S^{(2)} [x_0] = \frac{1}{4\pi\alpha'} \int \! d^2 & \sigma \,\Bigl\{ \nabla \ell^\rho \, \overline{\nabla} \ell^\sigma E_{\rho\sigma}
    + \ell^\rho \overline{\nabla} \ell^\sigma \, \CU_{\rho\sigma} + \ell^\rho {\nabla} \ell^\sigma \, \overline{\CU}_{\rho\sigma}
    + \! \ell^\rho \ell^\sigma \, \mathcal{V}_{\rho\sigma} \notag \\
    & + \lambda \ls \ell^\rho \, \overline{\nabla} \ell^\sigma \nabla_\rho \tau_\sigma \! + \tfrac{1}{2} \, \ell^\rho \ell^\sigma \, \overline{\p} x^\mu_0 \, \bigl( \nabla_\rho \nabla_\sigma \, \tau_\mu + \tau_\lambda \, R^\lambda{}_{\rho\sigma\mu} \bigr) \rs \notag \\[3pt]
    & + \overline{\lambda} \ls \ell^\rho \, {\nabla} \ell^\sigma \nabla_\rho \overline{\tau}_\sigma \! + \tfrac{1}{2} \, \ell^\rho \ell^\sigma \p x^\mu \, \bigl( \nabla_\rho \nabla_\sigma \, \overline{\tau}_\mu + \overline{\tau}_\lambda \, R^\lambda{}_{\rho\sigma\mu} \bigr) \rs \Bigr\}\,,
\end{align}
where
\begin{subequations} \label{eq:couplings}
\begin{align}
    \overline{\Gamma}_\rho & = \overline{\p} x^\mu_0\, E_{\mu\rho}\,,
        \qquad
    \Gamma_\rho = \p x^\mu_0\,E_{\mu\rho}\,,
        \qquad
    \Delta_\rho = \p x_0^\mu \, \overline{\p} x_0^\nu \, \bigl( \nabla_\rho E_{\mu\nu} + \mathcal{F}_{\rho\mu\nu} \bigr)\,, \\[6pt]
    \mathcal{U}_{\rho\sigma} & = \p x^\mu_0 \, \bigl( \nabla_{[\rho} E_{\sigma]\mu} - \tfrac{1}{2} \mathcal{F}_{\rho\sigma\mu} \bigr)\,, 
        \qquad\qquad\!%
    \overline{\CU}_{\rho\sigma} = \overline{\p} x^\mu_0 \, \bigl( \nabla_{[\rho} E_{\sigma]\mu} + \tfrac{1}{2} \mathcal{F}_{\rho\sigma\mu} \bigr)\,, \\[2pt]
      \mathcal{V}_{\rho\sigma} & = \frac{1}{2} \Bigl[ \p x^\mu_0 \, \overline{\p} x^\nu_0 \bigl( \nabla_{\!(\rho} \nabla_{\sigma)} E_{\mu\nu} - \nabla_\mu \nabla_{\!(\rho} \, E_{\sigma)\nu} - \nabla_\nu \nabla_{\!(\rho} E_{\sigma)\mu} + \nabla_{\!(\rho} \, \mathcal{F}_{\sigma)\mu\nu} \bigr)
    \notag \\
    & \hspace{2.6cm}
    + \p x_0^\mu \, \overline{\p} x_0^\nu \bigl( E_{\mu\lambda} R^\lambda{}_{(\rho\sigma)\nu} + E_{\nu\lambda} R^\lambda{}_{(\rho\sigma)\mu} \bigr) - 2 \overline{\nabla} {\p} x^\mu_0 \, \nabla_{(\rho} E_{\sigma)\mu} 
     \bigr) \Bigr]\,. \label{eq:GammaDeltaUV}
\end{align}
\end{subequations}
Here, $\mathcal{F}_{\rho\mu\nu}$ is the field strength for the anti-symmetric field $A_{\mu\nu}$\,, with
\be
    \mathcal{F}_{\rho\mu\nu} \equiv \p_\rho A_{\mu\nu} + \p_\mu A_{\nu\rho} + \p_\nu A_{\rho\mu}\,. 
\ee
Since we have shuffled all the $m_\mu^{A}$ dependence into $A_{\mu\nu}$ in \eqref{eq:eqsmaction}, it is understood that $m_\mu{}^A = 0$ in the connection coefficients $\Gamma^\rho{}_{\mu\nu}$ and hence the curvature tensor $R^\mu{}_{\nu\rho\sigma}$\,. It is understood that the coefficients in the expansion (such as the ones in \eqref{eq:couplings}) are evaluated at $x_0^\mu(\sigma)$\,.  
Moreover, we introduced the definitions
\be
    \nabla \equiv \nabla_1 -i \nabla_2\,, 
        \qquad
    \overline{\nabla} = \nabla_1 + i \nabla_2\,. 
\ee

The Vielbein postulates \eqref{eq:vielbeinpostulates} can be rewritten as
\be \label{eq:nablataunablaE}
    \nabla_\mu \tau_\nu{} = - \Omega_\mu \tau_\nu\,,
        \qquad
    \nabla_\mu \overline{\tau}_\nu = \Omega_\mu \overline{\tau}_\nu\,,
        \qquad
    \nabla_\mu E_\nu{}^{A'} = \Omega_\mu{}^{A'B'} E_\nu{}^{B'} - \Omega_\mu{}^{AA'} \tau_{\nu A}\,.
\ee
Using \eqref{eq:nablataunablaE} to rewrite $\nabla \tau$ in terms of $\Omega_\mu$ in \eqref{eq:S[x]}, and then applying the field redefinitions
\begin{subequations} \label{eq:redeflambda}
\begin{align}
    \lambda & \rightarrow \lambda \ls 1 + \ell^\rho \, \Omega_\rho + \tfrac{1}{2} \, \ell^\rho \ell^\sigma \bigl( \nabla_{\!\rho} \, \Omega_\sigma + \Omega_{\!\rho} \, \Omega_\sigma \bigr) + O (\ell^3) \rs, \\[4pt]
    \overline{\lambda} & \rightarrow \overline{\lambda} \ls 1 - \ell^\rho \, \Omega_\rho - \tfrac{1}{2} \, \ell^\rho \ell^\sigma \bigl( \nabla_{\!\rho} \Omega_\sigma - \Omega_{\!\rho} \, \Omega_\sigma \bigr) + O (\ell^3) \rs,
\end{align}
\end{subequations}
we find that \eqref{eq:S[x]} becomes
\begin{align} \label{eq:actioncovexpansion}
    S [x] = S [x_0] + & \frac{1}{4\pi\alpha'} \int d^2 \sigma \, \Bigl( \nabla \ell^\rho \, \overline{\Gamma}_\rho + \overline{\nabla} \ell^\rho \, \Gamma_\rho + \ell^\rho \, \Delta_\rho + \lambda \overline{\nabla} \ell^\rho \, \tau_\rho + \overline{\lambda} \, \nabla \ell^\rho \, \overline{\tau}_\rho \notag \\[2pt]
    & \hspace{1.2cm} + \nabla \ell^\rho \, \overline{\nabla} \ell^\sigma E_{\rho\sigma} + \ell^\rho \, \overline{\nabla} \ell^\sigma \, \CU_{\rho\sigma} + \ell^\rho \, {\nabla} \ell^\sigma \, \overline{\CU}_{\rho\sigma}
    + \ell^\rho \ell^\sigma \, \CV_{\rho\sigma} \notag \\[2pt]
    & \hspace{1.2cm} + \tfrac{1}{2} \, \lambda \, \ell^\rho \ell^\sigma \, \overline{\p} x^\mu_0 \, \tau_\lambda \, R^\lambda{}_{\rho\sigma\mu}
    + \tfrac{1}{2} \, \overline{\lambda} \, \ell^\rho \ell^\sigma \p x^\mu_0 \, \overline{\tau}_\lambda \, R^\lambda{}_{\rho\sigma\mu} \Bigr) + O (\ell^3)\,.
\end{align}
Note that the field redefinitions \eqref{eq:redeflambda} do not contribute any Jacobian in the path integral.

There is an alternative way to understand the set of field redefinitions in \eqref{eq:redeflambda} for the Lagrange multipliers. We initially assumed that $\lambda$ and $\overline{\lambda}$ were independent of the affine parameter $s$\,; we now assume that $\lambda$ and $\overline{\lambda}$ have the following $s$-dependence:
\be \label{eq:lambdas}
    \lambda (s) = \lambda_\mu (s) \, \tau^\mu [y (s)] \,,
        \qquad%
    \overline{\lambda} (s) = \overline{\lambda}_\mu (s) \, \overline{\tau}^\mu [y (s)] \,,
\ee
where 
\be
    \tau^\mu \equiv \frac{1}{2} \bigl( \tau^\mu{}_0 + \tau^\mu{}_1 \bigr)\,,
        \qquad
    \overline{\tau}^\mu \equiv \frac{1}{2} \bigl( \tau^\mu{}_0 - \tau^\mu{}_1 \bigr)\,,
\ee
and $\lambda_\mu$ and $\overline{\lambda}_\mu$ are required to satisfy
\be \label{eq:nablalambdamuconds}
    \nabla_{\!s} \, \lambda_\mu (s) = \nabla_{\!s} \, \overline{\lambda}_\mu (s) = 0\,,
\ee
together with
\be \label{eq:lambdamuconds}
    \lambda_\mu (s) = \lambda (s) \, \tau_\mu [y(s)]\,,
        \qquad
    \overline{\lambda}_\mu (s) = \overline{\lambda} (s)\, \overline{\tau}_\mu [y(s)]\,.
\ee
Here, \eqref{eq:lambdamuconds} is imposed such that no extra degree of freedom is introduced. 
It is natural to take the reparametrizations in \eqref{eq:lambdas} since $\lambda$ (and $\overline{\lambda}$) transforms with respect to the longitudinal Lorentz rotation in the same way as $\tau^\mu$ (and $\overline{\tau}^\mu$). Using the above prescriptions to expand $\lambda(s)$ and $\overline{\lambda}(s)$ as
\begin{subequations}
\begin{align}
    \lambda (1) & = \lambda (0) + \nabla_s \lambda (s) \big|_{s=0} + \tfrac{1}{2} \nabla_s^2 \lambda (s) \big|_{s=0} + \cdots\,, \\[4pt]
    \overline{\lambda} (1) & = \overline{\lambda} (0) + \nabla_s \overline{\lambda} (s) \big|_{s=0} + \tfrac{1}{2} \nabla_s^2 \overline{\lambda} (s) \big|_{s=0} + \cdots\,,
\end{align}
\end{subequations}
we obtain
\begin{subequations} \label{eq:lambda1overlinelambda1}
\begin{align}
    \lambda (1) & = \lambda (0) \ls 1 + \ell^\rho \Omega_\rho [x_0] + \tfrac{1}{2} \, \ell^\rho \ell^\sigma \bigl( \nabla_{\!\rho} \, \Omega_\sigma [x_0] + \Omega_{\!\rho} [x_0] \, \Omega_\sigma [x_0] \bigr) + \cdots \rs\,, \\[4pt]
    \overline{\lambda} (1) & = \overline{\lambda} (0) \ls 1 - \ell^\rho \Omega_\rho [x_0] - \tfrac{1}{2} \, \ell^\rho \ell^\sigma \bigl( \nabla_{\!\rho} \, \Omega_\sigma [x_0] - \Omega_{\!\rho} [x_0] \, \Omega_\sigma [x_0] \bigr) + \cdots \rs\,.
\end{align}
\end{subequations}
Plugging these expansions into \eqref{eq:Sys} has the same effect as applying the field redefinitions \eqref{eq:redeflambda} to \eqref{eq:S[x]}.

\section{Quantum Calculations} \label{sec:quancal}

In this section, we consider the path integral \eqref{eq:pathintegral} for the action \eqref{eq:actioncovexpansion} that is expanded around a covariant background. We have split the worldsheet field $x^\mu (\sigma)$ into the classical part $x_0^\mu (\sigma)$ and the quantum part parametrized by $\ell^\mu (\sigma)$\,.  For the additional worldsheet fields $\lambda (\sigma)$ and $\overline{\lambda} (\sigma)$\, in the sigma model, we split $\lambda (0)$ and $\overline{\lambda}(0)$ in \eqref{eq:lambda1overlinelambda1} as follows:
\be \label{eq:splitlambdarho}
    \lambda(0) = \lambda_0 (\sigma) + \rho (\sigma)\,,
        \qquad
    \overline{\lambda}(0) = \overline{\lambda}_0 (\sigma) + \overline{\rho} (\sigma)\,,
\ee
where $\lambda_0 (\sigma)$ and $\overline{\lambda}_0 (\sigma)$ are the classical parts on the same footing as $x_0^\mu (\sigma)$\,, while $\rho (\sigma)$ and $\overline{\rho} (\sigma)$ are quantum fields on the same footing as $\ell^\mu (\sigma)$\,.
There is an obvious choice for the measure in the path integral such that the path integral is covariant under the spacetime gauge transformations,
\be \label{eq:pathintegrallmu}
    \mathcal{Z} = \int \mathscr{D} \rho \, \mathscr{D} \, \overline{\rho} \, \mathscr{D} \ell^\mu \sqrt{G[x_0]} \, e^{- S[\rho\,, \, \overline{\rho}\,, \, \ell^\mu]}\,. 
\ee
Here, the action is given in \eqref{eq:actioncovexpansion}, and $G[x_0]$ is defined in \eqref{eq:det} but now evaluated at $x_0$\,. For later convenience, we change variables from $\ell^\mu$ to $\ell^I$, $I \in \{A, A'\}$\,, with
\be
    \ell^A \equiv \tau_\mu{}^A \, \ell^\mu\,,
        \qquad
    \ell^{A'} \equiv E_\mu{}^{A'} \ell^\mu\,.
\ee
The path integral in \eqref{eq:pathintegrallmu} now becomes
\be \label{eq:pathintegrallAlA'}
    \mathcal{Z} = \int \mathscr{D} \rho \, \mathscr{D} \, \overline{\rho} \, \mathscr{D} \ell^I \exp \Bigl\{ - S\bigl[ \rho\,, \, \overline{\rho} \,, \ell^I \bigr] \Bigr\}\,,
\ee
where the measure takes a more canonical form, which can be defined with respect to the $L^2$ norms \cite{Deligne:1999qp}
\be
    \bigl|\!\bigl| \ell^A \bigr|\!\bigr|^2 = \int d^2 \sigma \, \ell^A \ell^B \eta_{AB}\,,
        \qquad
    \bigl|\!\bigl| \ell^{A'} \bigr|\!\bigr|^2 = \int d^2 \sigma \, \ell^{A'} \ell^{B'} \delta_{A'B'}\,.
\ee
Since the second norm breaks the boost symmetry, the full string Newton-Cartan gauge symmetry will only become manifest at the end of the calculation. 
The action in \eqref{eq:pathintegrallAlA'} is the same as \eqref{eq:actioncovexpansion} but with the substitutions \eqref{eq:splitlambdarho} and $\ell^\mu = \tau^\mu{}_{\!A} [x_0] \, \ell^A + E^\mu{}_{\!A'} [x_0] \, \ell^{A'}$\,, where it is useful to apply \eqref{eq:nablataunablaE} to derive
\be
    \nabla_\alpha \ell^\mu = \tau^\mu{}_{\!A}[x_0] \, D_\alpha \ell^A + E^\mu{}_{\!A'}[x_0] \, D_\alpha \ell^{A'}\,, 
\ee
with
\begin{subequations}
\begin{align}
    D_\alpha \ell^A & \equiv \p_\alpha \ell^A - \epsilon^A{}_B \, \Omega_\mu[x_0] \, \p_\alpha x^\mu_0 \, \ell^B\,, \\[2pt] 
    D_\alpha \ell^{A'} & \equiv \p_\alpha \ell^{A'} + \bigl( \eta_{AB} \, \Omega_{\mu}{}^{AA'}[x_0] \, \ell^B - \Omega_\mu{}^{A'B'} [x_0] \, \ell^{B'} \bigr) \, \p_\alpha x^\mu_0 \,.
\end{align}
\end{subequations}

Next, we collect terms quadratic in the quantum fields $\{ \rho\,, \overline{\rho}\,, \ell^I\}$\,. These terms contribute the one-loop effective action. The free part of the quadratic action is
\begin{align} \label{eq:Sfree}
    S_\text{free} = \frac{1}{4\pi\alpha'} \int d^2 \sigma \bigl( \p \ell^{A'} \, \overline{\p} \ell^{A'} + \rho \, \overline{\p} \ell + \overline{\rho} \, \p \overline{\ell} \, \bigr)\,,
\end{align}
with $\ell \equiv \ell^0 + \ell^1$ and $\overline{\ell} \equiv \ell^0 - \ell^1$\,. The interactions between the quantum fields and the background fields are
\begin{align} \label{eq:Sint(2)}
     S_{\text{int}}^{(2)} = \frac{1}{4 \pi \alpha'} \int d^2 \sigma \, \Bigl[ \ell^I \, \overline{\p} \ell^J \mathcal{A}_{IJ} [x_0] + \ell^I \, \p \ell^J \bar{\mathcal{A}}_{IJ} [x_0] & + \ell^I \ell^J \mathcal{B}_{IJ} [x_0] \notag \\
    & + \rho \, \ell \, \overline{\mathcal{C}} [x_0] + \overline{\rho} \, \overline{\ell} \, \mathcal{C} [x_0] \Bigr]\,,
\end{align}
where components of the background dependent coefficients $\CA_{IJ}[x_0]$\,, $\bar{\CA}_{IJ} [x_0]$\,, $\CB_{IJ} [x_0]$\,, $\CC [x_0]$ and $\overline{\CC} [x_0]$ can be read off from \eqref{eq:actioncovexpansion}. Note that $\mathcal{A}_{IJ}$ and $\bar{\CA}_{IJ}$ are anti-symmetric while $\CB_{IJ}$ is symmetric. Later in this section, when we compute the one-loop effective action, we will find that only the following components of these background-dependent coefficients contribute:
\begin{subequations} \label{eq:ABCABC}
\begin{align}
    \mathcal{A}_{A' B'} & = \Omega_{\mu}{}^{A' B'} \partial x^\mu_0 +\mathcal{U}_{A' B'}\,, 
        \qquad\quad
    \mathcal{C} = - \Omega_\mu \, \p x^\mu_0\,,  \\[4pt]
    \bar{\mathcal{A}}_{A' B' } & = \Omega_{\mu}{}^{A' B'} \, \overline{\partial} x^\mu_0 + \overline{\mathcal{U}}_{A' B'}\,,
        \qquad\quad\!
    \overline{\mathcal{C}} = \Omega_\mu \, \overline{\p} x^\mu_0\,, \\[2pt]
    \mathcal{B}_{A' B' }& = 
    \Omega_{\mu}{}^{(A'}{}_{C'} \Big( \Omega_{\nu}{}^{B') C'} \partial x^\mu_0 \, \overline{\partial} x^\nu_0 + \CU^{B')C'} \, \overline{\p} x^\mu_0 + \overline{\CU}{}^{B')C'} \, \p x^\mu_0 \Bigr) \notag \\[2pt]
    & \hspace{3cm} + \mathcal{V}_{A' B'} + \tfrac{1}{2} \, E^\rho{}_{\!A'} E^\sigma{}_{\!B'} \bigl( \lambda_0 \, \overline{\p} x_0^\mu \, \tau_\nu + \overline{\lambda}_0 \, \p x_0^\mu \, \overline{\tau}_\nu \bigr) R^\nu{}_{\!(\rho\sigma)\mu}\,.
\end{align}
\end{subequations}
Here, $\CU_{A'B'} \equiv E^\mu{}_{\!A'} E^\nu{}_{\!B'} \, \CU_{\mu\nu}$ and $\CV_{A'B'} \equiv E^\mu{}_{\!A'} E^\nu{}_{\!B'} \, \CV_{\mu\nu}$\,, with $\CU_{\mu\nu}$ and $\CV_{\mu\nu}$ given in \eqref{eq:couplings}. 
Since the explicit expressions of other components in the background dependent coefficients are not needed for later calculations, we will not present them here. 

\subsection{One-loop quantum corrections}\label{sectionwithFeynmanrules}

We now derive the Feynman rules associated with \eqref{eq:Sfree} and \eqref{eq:Sint(2)}, and then compute one-loop diagrams to determine the one-loop effective action. 

We start with defining the expectation value for a given operator $\mathcal{O}$ with respect to the free theory \eqref{eq:Sfree} as
\be
    \langle 0| \mathcal{O} |0 \rangle \equiv \int \mathscr{D} \rho \, \mathscr{D} \overline{\rho} \, \mathscr{D} \mathscr{\ell}^I \, \mathcal{O} \, e^{- S_\text{free}}\,.
\ee
Here, we take the normalization such that $\langle 0 | 0 \rangle = 1$\,. 
From the free action \eqref{eq:Sfree} we derive all the nontrivial propagators in position space,
\begin{subequations}
\begin{align}
	G^{A'B'} (\sigma\,, \sigma') & \equiv \langle 0 | \CT \, \ell^{A'} \! (\sigma) \, \ell^{B'} \! (\sigma') | 0 \rangle
=
2 \pi \alpha' \, \delta^{A'B'} \Delta (\sigma - \sigma')\,, \label{eq:GA'B'} \\[4pt]
	G (\sigma\,, \sigma')& \equiv \langle 0| \CT \, \rho (\sigma) \, \ell (\sigma') |0 \rangle
=
4 \pi \alpha' \, \p \Delta (\sigma - \sigma')\,, \label{eq:Grhoell} \\[4pt]
\overline{G} (\sigma\,, \sigma')& \equiv \langle 0 | \CT \, \overline{\rho} (\sigma) \, \overline{\ell} (\sigma') | 0 \rangle
=
4 \pi \alpha' \, \overline{\p} \Delta (\sigma - \sigma')\,,
\end{align}
\end{subequations}
where $\p$ and $\overline{\p}$ are with respect to $\sigma$ (later, we will also use $\p'$ and $\overline{\p}'$ to denote derivatives with respect to $\sigma'$) and
\be
    \Delta (\sigma - \sigma') \equiv \int \frac{d^2 k}{(2\pi)^2} \frac{e^{i k \cdot ( \sigma - \sigma' )}}{k^2 - i \epsilon} = - \frac{1}{2\pi} \! \ls \log \bigl( \tfrac{1}{2} \mu^{}_\text{IR} |\sigma - \sigma'| \bigr) + \gamma^{}_\text{E} \rs + O (\mu^{}_\text{IR})\,,
\ee
with $\mu^{}_\text{IR}$ an infrared (IR) regulator and $\gamma^{}_\text{E}$ the Euler-Mascheroni constant.

To present the Feynman rules for vertices, we define
\be
    d\mu (\sigma\,, \sigma'\,, \sigma'') \equiv d^2 \sigma'' \, \delta^{(2)} (\sigma'' - \sigma) \, \delta^{(2)} (\sigma'' - \sigma')\,. 
\ee
The vertices associated with the operators in \eqref{eq:Sint(2)} are
\begin{subequations}
\begin{align}
\begin{minipage}{4cm}
\begin{tikzpicture}
    \draw (0,0) -- (1.1,0) (1.3,0) -- (2.4,0);
    \draw (1.2,0) circle (1mm);
    \node at (1.2,0) {\scalebox{0.75}{$\times$}};
    \node at (1.2,0.35) {$\mathcal{A}$};
    \node at (0,-0.35) {$\ell^I (\sigma)$};
    \node at (2.4,-0.35) {$\ell^J (\sigma')$};
\end{tikzpicture}
\end{minipage}
\hspace{-5mm}
& \equiv V^\CA_{IJ} (\sigma\,, \sigma') = - \frac{1}{4\pi\alpha'} \int \CA_{IJ} (\sigma'') \bigl( \overline{\p}' - \overline{\p} \bigr) d\mu (\sigma\,, \sigma'\,, \sigma'')\,, \\
\begin{minipage}{4cm}
\begin{tikzpicture}
    \draw (0,0) -- (1.1,0) (1.3,0) -- (2.4,0);
    \draw (1.2,0) circle (1mm);
    \node at (1.2,0) {\scalebox{0.75}{$\times$}};
    \node at (1.2,0.35) {$\bar{\CA}$};
    \node at (0,-0.35) {$\ell^I (\sigma)$};
    \node at (2.4,-0.35) {$\ell^J (\sigma')$};
\end{tikzpicture}
\end{minipage}
\hspace{-5mm}
& \equiv V^{\bar{\CA}}_{IJ} (\sigma\,, \sigma') = - \frac{1}{4\pi\alpha'} \int \bar{\CA}_{IJ} (\sigma'') \bigl( {\p}' - {\p} \bigr) d\mu (\sigma\,, \sigma'\,, \sigma'')\,, \\
\begin{minipage}{4cm}
\begin{tikzpicture}
    \draw (0,0) -- (1.1,0) (1.3,0) -- (2.4,0);
    \draw (1.2,0) circle (1mm);
    \node at (1.2,0) {\scalebox{0.75}{$\times$}};
    \node at (1.2,0.35) {$\mathcal{B}$};
    \node at (0,-0.35) {$\ell^I (\sigma)$};
    \node at (2.4,-0.35) {$\ell^J (\sigma')$};
\end{tikzpicture}
\end{minipage}
\hspace{-5mm}
& \equiv V^\CB_{IJ} (\sigma\,, \sigma') = - \frac{1}{2\pi\alpha'} \int \CB_{IJ} (\sigma'') \, d\mu (\sigma\,, \sigma'\,, \sigma'')\,, \\
\begin{minipage}{4cm}
\begin{tikzpicture}
    \draw (0,0) -- (1.1,0) (1.3,0) -- (2.4,0);
    \draw (1.2,0) circle (1mm);
    \node at (1.2,0) {\scalebox{0.75}{$\times$}};
    \node at (1.2,0.35) {$\mathcal{C}$};
    \node at (0,-0.35) {$\rho (\sigma)$};
    \node at (2.4,-0.35) {$\ell (\sigma')$};
\end{tikzpicture}
\end{minipage}
\hspace{-5mm} 
& \equiv V^\CC (\sigma\,, \sigma') = - \frac{1}{4\pi\alpha'} \int \CC (\sigma'') \, d\mu (\sigma\,, \sigma'\,, \sigma'')\,, \\
\begin{minipage}{3.5cm}
\begin{tikzpicture}
    \draw (0,0) -- (1.1,0) (1.3,0) -- (2.4,0);
    \draw (1.2,0) circle (1mm);
    \node at (1.2,0) {\scalebox{0.75}{$\times$}};
    \node at (1.2,0.35) {$\overline{\mathcal{C}}$};
    \node at (0,-0.35) {$\overline{\rho} (\sigma)$};
    \node at (2.4,-0.35) {$\overline{\ell} (\sigma')$};
\end{tikzpicture}
\end{minipage}
& \equiv V^{\overline{\CC}} (\sigma\,, \sigma') = - \frac{1}{4\pi\alpha'} \int \overline{\CC} (\sigma'') \, d\mu (\sigma\,, \sigma'\,, \sigma'')
\end{align}
\end{subequations}

We first consider one-loop diagrams with a single vertex insertion. Since $\mathcal{A}_{IJ}$ and $\bar{\CA}_{IJ}$ are anti-symmetric, we have
\begin{align}
\begin{minipage}{3cm}
\begin{tikzpicture}
    \draw (0,0) .. controls (-1.5,1.5) and (1.5,1.5) .. (0,0);
    \filldraw [white] (0,0) circle [radius=0.115];
    \draw (0,0) circle [radius=0.115];
    \node at (0,0) {\scalebox{0.86}{$\times$}};
    \node at (0,-0.35) {$\mathcal{A}$};
\end{tikzpicture}
\end{minipage}
\hspace{-8mm}
=
\hspace{-8mm}
\begin{minipage}{3cm}
\begin{tikzpicture}
    \draw (0,0) .. controls (-1.5,1.5) and (1.5,1.5) .. (0,0);
    \filldraw [white] (0,0) circle [radius=0.115];
    \draw (0,0) circle [radius=0.115];
    \node at (0,0) {\scalebox{0.86}{$\times$}};
    \node at (0,-0.35) {$\bar{\mathcal{A}}$};
\end{tikzpicture}
\end{minipage}
\hspace{-8mm}
= 0\,.
\end{align}
The remaining one-loop diagrams with a single vertex insertion sum to be
\vspace{-3mm}
\begin{align}
&
\begin{minipage}{3cm}
\begin{tikzpicture}
    \draw (0,0) .. controls (-1.5,1.5) and (1.5,1.5) .. (0,0);
    \filldraw [white] (0,0) circle [radius=0.115];
    \draw (0,0) circle [radius=0.115];
    \node at (0,0) {\scalebox{0.86}{$\times$}};
    \node at (0,-0.35) {$\mathcal{B}$};
\end{tikzpicture}
\end{minipage}
\hspace{-8mm}
+
\hspace{-8mm}
\begin{minipage}{3cm}
\begin{tikzpicture}
    \draw (0,0) .. controls (-1.5,1.5) and (1.5,1.5) .. (0,0);
    \filldraw [white] (0,0) circle [radius=0.115];
    \draw (0,0) circle [radius=0.115];
    \node at (0,0) {\scalebox{0.86}{$\times$}};
    \node at (0,-0.35) {${\mathcal{C}}$};
\end{tikzpicture}
\end{minipage}
\hspace{-8mm}
+
\hspace{-8mm}
\begin{minipage}{3cm}
\begin{tikzpicture}
    \draw (0,0) .. controls (-1.5,1.5) and (1.5,1.5) .. (0,0);
    \filldraw [white] (0,0) circle [radius=0.115];
    \draw (0,0) circle [radius=0.115];
    \node at (0,0) {\scalebox{0.86}{$\times$}};
    \node at (0,-0.35) {$\overline{\mathcal{C}}$};
\end{tikzpicture}
\end{minipage}
\hspace{-8mm} \notag 
\end{align}
\vspace{-6mm}
\begin{align} \label{eq:lightball0}
& = \frac{1}{2} \int d^2 \sigma \, d^2 \sigma' \ls V^\CB_{A'B'} (\sigma\,, \sigma') \, G^{A'B'} (\sigma\,, \sigma') + V^\CC (\sigma\,, \sigma') \, \overline{G} (\sigma\,, \sigma') + V^{\overline{\CC}} (\sigma\,, \sigma') \, {G} (\sigma\,, \sigma') \rs \notag \\
& = - \frac{1}{2} \, \Delta (0) \int d^2 \sigma \, \CB^{A'}{}_{\!\!A'} \notag \\
& \quad - \frac{1}{2} \int d^2 \sigma \Bigl\{ \CC(\sigma) \ls \overline{\p} \Delta (\sigma - \sigma') \rs_{\sigma' = \sigma} + \overline{\mathcal{C}} (\sigma) \ls \p \Delta (\sigma - \sigma') \rs_{\sigma' = \sigma} \Bigr\}\,.
\end{align}
Note the log divergent factor
\be \label{eq:Delta(0)}
    \Delta (0) = \frac{1}{2\pi} \log \! \lr \frac{\Lambda}{\mu^{}_\text{IR}} \rr ,
\ee
where we have introduced the UV cutoff $\Lambda$ in momentum space. 
In addition,
\be \label{eq:Ckappa}
    \int d^2 \sigma \, \CC(\sigma) \! \ls \overline{\p} \Delta (\sigma - \sigma') \rs_{\sigma' = \sigma} = \int \! d^2 \sigma \, \CC(\sigma) \int \frac{d^2k}{(2\pi)} \frac{i}{\kappa}\,,
\ee
where $\kappa \equiv k_1 - i k_2$\,. The momentum integral in \eqref{eq:Ckappa} is linearly divergent and needs to be treated with care.  We will see later that taking the na\"{i}ve regularization by setting $\ls \overline{\p} \Delta (\sigma - \sigma') \rs_{\sigma' = \sigma} = 0$ turns the string-Galilean boost symmetry anomalous in the final beta-functions. However, a linear divergence is defined up to a boundary term \cite{Zee:2003mt}, which can be fixed in this case by requiring the boost symmetry. Consider a shift in $\kappa$\,, parametrized by an arbitrary constant $a$\,, such that
\begin{align} \label{eq:inteCdDelta}
    \int \! d^2 \sigma \, \CC(\sigma) \! \ls \overline{\p} \Delta (\sigma - \sigma') \rs_{\sigma' = \sigma}
    & = \int \! d^2 \sigma \int \! \frac{d^2 q}{(2\pi)^2} \, \tilde{\CC} (q) \, e^{i q \cdot \sigma} \int \frac{d^2 k}{(2\pi)^2} \frac{i}{\kappa + a \, (q_1 - i q_2)} + \cdots \notag \\
    & = - a \, \Delta (0) \int \! d^2 \sigma \int \! \frac{d^2 q}{(2\pi)^2} \, i \bigl( q_1 + i q_2 \bigr) \, \tilde{\CC} (q) \, e^{i q \cdot \sigma} + \cdots \notag \\
    & = - a \, \Delta (0) \int d^2 \sigma \, \overline{\p} \, \CC + \cdots\,.
\end{align}
Here, $\tilde{\CC} (q)$ is the Fourier transform of $\CC (\sigma)$ and ``$\cdots$" denotes finite contributions. We have regularized all linear divergences to zero but only kept the boundary contributions, which will also contribute a boundary term to the one-loop effective action. 
Similarly, we take
\begin{align} \label{eq:inteCdDelta2}
    \int \! d^2 \sigma \, \overline{\CC}(\sigma) \! \ls {\p} \Delta (\sigma - \sigma') \rs_{\sigma' = \sigma}
    & = - a \, \Delta (0) \int d^2 \sigma \, {\p} \, \overline{\CC} + \cdots\,.
\end{align}
We will later see that requiring the boost symmetry in the resulting beta-functions fixes $a = 1$\,. Keeping the parameter $a$ in \eqref{eq:inteCdDelta} and \eqref{eq:inteCdDelta2}, we find that \eqref{eq:lightball0} becomes
\vspace{-3mm}
\begin{align} \label{eq:lightball}
&
\begin{minipage}{3cm}
\begin{tikzpicture}
    \draw (0,0) .. controls (-1.5,1.5) and (1.5,1.5) .. (0,0);
    \filldraw [white] (0,0) circle [radius=0.115];
    \draw (0,0) circle [radius=0.115];
    \node at (0,0) {\scalebox{0.86}{$\times$}};
    \node at (0,-0.35) {$\mathcal{B}$};
\end{tikzpicture}
\end{minipage}
\hspace{-8mm}
+
\hspace{-8mm}
\begin{minipage}{3cm}
\begin{tikzpicture}
    \draw (0,0) .. controls (-1.5,1.5) and (1.5,1.5) .. (0,0);
    \filldraw [white] (0,0) circle [radius=0.115];
    \draw (0,0) circle [radius=0.115];
    \node at (0,0) {\scalebox{0.86}{$\times$}};
    \node at (0,-0.35) {${\mathcal{C}}$};
\end{tikzpicture}
\end{minipage}
\hspace{-8mm}
+
\hspace{-8mm}
\begin{minipage}{3cm}
\begin{tikzpicture}
    \draw (0,0) .. controls (-1.5,1.5) and (1.5,1.5) .. (0,0);
    \filldraw [white] (0,0) circle [radius=0.115];
    \draw (0,0) circle [radius=0.115];
    \node at (0,0) {\scalebox{0.86}{$\times$}};
    \node at (0,-0.35) {$\overline{\mathcal{C}}$};
\end{tikzpicture}
\end{minipage}
\hspace{-8mm} 
= - \frac{1}{2} \, \Delta (0) \int d^2 \sigma \, \ls \CB^{A'}{}_{\!\!A'} - a \, \bigl( \overline{\p} \mathcal{C} + \p \overline{\CC} \, \bigr) \, \rs.
\end{align}
\vspace{-3mm}

We now consider one-loop diagrams with two vertex insertions. There is only one non-vanishing contribution,
$$
\begin{tikzpicture}
\draw (0,0) to [out=90,in=90] (2,0);
\draw (0,0) to [out=-90,in=-90] (2,0);
\filldraw [white] (0,0) circle [radius=0.115];
\draw (0,0) circle [radius=0.115];
\node at (0,0) {\scalebox{0.85}{$\times$}};
\filldraw [white] (2,0) circle [radius=0.115];
\draw (2,0) circle [radius=0.115];
\node at (2,0) {\scalebox{0.85}{$\times$}};
\node at (-0.35,0) {$\CA$};
\node at (2.35,0) {$\bar{\CA}$};
\end{tikzpicture} 
$$
\vspace{-1cm}
\begin{align} \label{eq:AAbardiagram}
& = \frac{1}{2} \int d^2\sigma \, d^2 \sigma' \, d^2 \tilde{\sigma} \, d^2 \tilde{\sigma}' \, V^\CA_{A'B'} (\sigma\,, \tilde{\sigma}) \, G^{B'C'} (\tilde{\sigma}\,, \sigma') \,  V^{\bar{\CA}}_{C'D'} (\sigma'\,, \tilde{\sigma}') \, G^{D'A'} (\tilde{\sigma}'\,, \sigma) \notag \\
& = \frac{1}{2} \, \Delta(0) \int d^2 \sigma \, \CA^{A'B'} \, \bar{\CA}_{A'B'} + \text{finite}\,.
\end{align}
Other diagrams of the same type contain either $\p x^\mu \, \p x^\nu$ or $\overline{\p} x^\mu \, \overline{\p} x^\nu$ and they vanish identically by rotational symmetry on the worldsheet. 
Summing over \eqref{eq:lightball} and \eqref{eq:AAbardiagram}, we obtain the one-loop contribution to the effective action,
\be \label{eq:Seff0}
    S_\text{1-loop} = \frac{1}{2} \, \Delta (0) \int d^2 \sigma \ls - \CA^{A'B'} \bar{\CA}_{A'B'} + \CB^{A'}{}_{\!\!A'} - a \, \bigl( \overline{\p} \CC + \p \overline{\CC} \, \bigr) \rs,
\ee
where we only kept the log divergent terms.

Plugging \eqref{eq:ABCABC} in \eqref{eq:Seff0}, and then using \eqref{eq:couplings} to substitute $\CU$ and $\CV$ gives
\begin{align} \label{eq:Seffexp}
    S_\text{1-loop} = \frac{1}{2} \, \Delta (0) \! \int d^2 \sigma \, \Bigl[ \p x^\mu_0 \, \overline{\p} x^\nu_0 \, \CO_{\mu\nu} & - E^{\rho\sigma} \, \overline{\nabla} \p x^\mu_0 \, \nabla_{\rho} E_{\sigma\mu} \notag \\
    & + \tfrac{1}{2} E^{\rho\sigma} \bigl( \lambda_0 \, \overline{\p} x_0^\mu \, \tau_\lambda + \overline{\lambda}_0 \, \p x_0^\mu \, \overline{\tau}_\lambda \bigr) R^\lambda{}_{\rho\sigma\mu} \Bigr]\,,
\end{align}
where
\begin{align}
    \CO_{\mu\nu} & = \tfrac{1}{2} E^{\rho\sigma} \bigl( \nabla_{\!\rho} \nabla_{\!\sigma} E_{\mu\nu} - \nabla_{\!\mu} \nabla_{\!\rho} E_{\sigma\nu} - \nabla_\nu \nabla_{\!\rho} E_{\sigma \mu} + \nabla_{\!\rho} \, \CF_{\sigma\mu\nu} \bigr) \notag \\[3pt]
    & \quad - 2 \, a \, \p_{[\mu} \Omega_{\nu]} + \tfrac{1}{2} E^{\rho\sigma}\bigl( E_{\mu\lambda} R^\lambda{}_{\rho\sigma\nu} + E_{\nu\lambda} R^\lambda{}_{\rho\sigma\mu} \bigr) + \tfrac{1}{4} E^{\rho\sigma} E^{\kappa\lambda} \CF_{\rho\kappa\mu} \, \CF_{\sigma\lambda\nu} \notag \\[4pt]
    & \quad  - E^{\rho\sigma} E^{\kappa\lambda} \, \nabla_{\![\rho} E_{\kappa]\mu} \nabla_{[\sigma} E_{\lambda]\nu} 
    + \tfrac{1}{2} E^{\rho\sigma} E^{\kappa\lambda} \bigl( \CF_{\rho\kappa\mu} \nabla_\sigma E_{\lambda\nu} - \CF_{\rho\kappa\nu} \nabla_\sigma E_{\lambda\mu} \bigr)\,.
\end{align}
Note that the $(\lambda_0\,, \overline{\lambda}_0)$-dependent terms in \eqref{eq:Seffexp} vanish identically since 
\be \label{eq:EtauR0}
    E^{\rho\sigma} \tau_\nu{}^A R^\nu{}_{\rho\sigma\mu} = 0\,.
\ee
This identity is manifestly true by using the following relation from \cite{stringyNClimit} but with $m_\mu{}^A$ set to zero, 
\begin{align} \label{eq:curvaturetensor2form}
    R^\mu{}_{\nu\rho\sigma} = - \epsilon^A{}_B \, \tau^\mu{}_{\!A} \, \tau_\nu{}^{B} \, \CR_{\rho\sigma} (M) & + E^\mu{}_{\!A'} \, \tau_{\nu A} \, \CR_{\rho\sigma}{}^{\!AA'} \! (G)
        - E^\mu{}_{\!A'} \, E_\nu{}^{B'} \, \CR_{\rho\sigma}{}^{\!A'B'} \! (J)\,,
\end{align}
where $\CR_{\mu\nu} (M)\,, \CR_{\mu\nu}{}^{AA'} (G)$ and $\CR_{\mu\nu}{}^{A'B'} (J)$ are curvature two-forms associated with the generators $M$ for longitudinal Lorentz rotation, $G_{AA'}$ for string-Galilean boost and $J_{A'B'}$ for transverse rotation. In particular, note that
\be \label{eq:dOmega}
    \CR_{\mu\nu} (M) \equiv 2 \, \p_{[\mu} \Omega_{\nu]}\,.
\ee

It is convenient to further rewrite \eqref{eq:Seffexp} by introducing the following shorthand notation: we exchange a lowered curved index $\mu$ with flat indices in $\{A\,, A'\}$ when it is contracted with an inverse Vielbein field and there is no derivative acting on this inverse Vielbein field. For example, for a tensor $\CT_{\mu\nu}$ contracted with the Vielbein fields $E^\mu{}_{A'}$ and $\tau^\mu{}_{A}$\,, we have
\be \label{eq:exsimnotation}
    \CT_{A'A} = E^{\mu}{}_{\!A'} \, \tau^\nu{}_{\!A} \, \CT_{\mu\nu}\,.
\ee
In this notation, using \eqref{eq:EtauR0}, \eqref{eq:curvaturetensor2form} and \eqref{eq:dOmega}, \eqref{eq:Seffexp} can be rewritten as 
\be \label{eq:ShortenSeff}
    S_\text{1-loop} = \frac{1}{2} \Delta(0) \int d^2 \sigma \lr \p x^\mu_0 \, \overline{\p} x^\nu_0 \, \CO_{\mu\nu} - \overline{\nabla} \p x_0^\mu \, \nabla^{A'} \! E_{A'\mu} \rr\,,
\ee
where
\begin{align}
    \CO_{\mu\nu} & = \tfrac{1}{2} \bigl( \nabla^{\!A'} \nabla_{\!A'} E_{\mu\nu} - \nabla_{\!\mu} \nabla^{\!A'} E_{A' \nu} - \nabla_{\!\nu} \nabla^{\!A'} E_{A' \mu} + \nabla^{\!A'} \CF_{\!A' \mu\nu} \bigr) \notag \\[3pt]
    & \quad - a \, \CR_{\mu\nu} (M) - E_{(\mu}{}^{B'} \CR^{}_{\nu)A'}{}^{\!A'B'} (J)  + \tfrac{1}{4} \, \CF_{A'B'\mu} \, \CF_{A'B'\nu} \notag \\[4pt]
    & \quad - \nabla_{\![A'} E_{B']\mu} \nabla_{[A'} E_{B']\nu} 
    + E^{\rho\sigma} E^{\kappa\lambda} \CF_{A'B'[\mu} \nabla^{[A'} E^{B']}{}_{\nu]}\,.
\end{align}

\subsection{Quantum holomorphic and anti-holomorphic conditions} \label{sec:hantihcond}

To proceed with deriving the beta-functions from the one-loop corrections \eqref{eq:ShortenSeff} to the effective action, we need to note a few path integral identities that generalize the classical holomorphic and anti-holomorphic conditions imposed by integrating out the Lagrange multipliers $\lambda$ and $\overline{\lambda}$ in the sigma model action \eqref{eq:eqsmaction} in the flat limit. These identities will prove to be essential in later discussions.

In the flat limit $E_\mu{}^{A'} \rightarrow \delta_\mu^{A'}\,, \tau_\mu{}^A = \delta_\mu^A$ and $m_\mu{}^A \rightarrow 0$\,, without any $B$-field or dilaton, the sigma model action \eqref{eq:eqsmaction} becomes
\be
    S_\text{free} [\lambda\,, \overline{\lambda}\,, x^\mu] = \frac{1}{4\pi\alpha'} \int d^2 \sigma \lr \p x^{A'} \, \overline{\p} x^{A'} + \lambda \, \overline{\p} X + \overline{\lambda} \, \p \overline{X} \rr,
\ee
where $X \equiv x^0 + x^1$ and $\overline{X} \equiv x^0 - x^1$\,.
Here, the worldsheet fields $\lambda$ and $\overline{\lambda}$ play the role of a Lagrange multiplier that impose, respectively, the holomorphic and anti-holomorphic conditions $\overline{\p} X = \p \overline{X} = 0$\,.
On a curved background, with respect to the classical background field $x_0^\mu (\sigma)$\,, the analogue of these (anti)-holomorphic conditions are
\be
    \overline{D} X_0 = D \overline{X}_0 = 0\,,
\ee
where we introduced the notation
\be \label{eq:dX0dx0A'}
    D_\alpha X_0 \equiv \p_\alpha x^\mu_0 \, \tau_\mu [x_0]\,,
        \quad
    D_\alpha \overline{X}_0 \equiv \p_\alpha x^\mu_0 \, \overline{\tau}_\mu [x_0]\,,
        \quad
    D_\alpha x^{A'}_0 \equiv \p_\alpha x^\mu_0 \, E_\mu{}^{A'} [x_0]\,,
\ee
In the following, we discuss implications of these classical constraints for various correlation functions.

Consider correlation functions evaluated with respect to the background field $x_0^\mu\,, \lambda_0$ and $\overline{\lambda}_0$\,. For any operator $\mathcal{O}$\,, define
\begin{align} \label{eq:Ox0}
    \langle \mathcal{O} \rangle_{0} & \equiv \int \mathscr{D} \lambda_0 \, \mathscr{D} \overline{\lambda}_0 \, \mathscr{D} x_0^\mu \sqrt{G[x_0]} \, \mathcal{O} \, e^{-S [\lambda_0, \overline{\lambda}_0, x_0]}\,,
\end{align}
with
\begin{align} \label{eq:S00}
    S[\lambda_0\,, \overline{\lambda}_0\,, x_0] & = \frac{1}{4\pi\alpha'} \int d^2 \sigma \Bigl[ \p x^\mu_0 \, \overline{\p} x^\nu_0 \, \bigl( E_{\mu\nu} + A_{\mu\nu} \bigr) + \lambda_0 \overline{D} X_0 + \overline{\lambda}_0 D \overline{X}_0 \Bigr]\,. 
\end{align}
Here, we also take $\lambda_0\,, \overline{\lambda}_0$ and $x_0$ to be quantum fields and they are integrated over in the path integral, which was not necessary when we only integrated over $\rho\,, \overline{\rho}$ and $\ell$ to derive the one-loop effective action.\,\footnote{Note that we already integrated out $\rho\,, \overline{\rho}$ and $\ell$ at this point.} 

We first consider the correlation function
\begin{align*}
    & \quad \Bigl\langle \bigl[ \lambda_0 (\sigma) \bigr]^n \, \bigl[ \overline{D} X_0 (\sigma) \bigr]^{\!n+1} \, \mathcal{O} \bigl[ x_0\,, \overline{\lambda}_0 \bigr](\sigma) \Bigr\rangle_{\!0} \notag \\[2pt]
    & = \int \! \mathscr{D} \lambda_0 \, \mathscr{D} \overline{\lambda}_0 \, \mathscr{D} x_0 \sqrt{G [x_0]} \, \, \frac{\delta}{\delta \lambda_0 (\sigma')} \Bigl( M\bigl[\lambda_0\,,x_0\bigr] (\sigma\,, \sigma') \, \mathcal{O} [\overline{\lambda}_0\,, x_0] (\sigma) \, e^{-S[x_0]} \Bigr) \bigg|_{\sigma'=\sigma}\,,   
\end{align*}
where $\mathcal{O} \bigl[ \overline{\lambda}_0\,, x_0 \bigr]$ can be any operator insertion and
\begin{align}
    M [\lambda_0\,, x_0] (\sigma, \sigma') & = - n! \sum_{r=1}^{n+1} \frac{(4\pi\alpha')^r \ls \delta^{(2)} (\sigma - \sigma') \rs^{r-1}}{(n+1-r)!} \, \ls \lambda_0 (\sigma) \, \overline{D} X_0 (\sigma) \rs^{n+1-r}\,.
\end{align}
Since a total functional derivative term under the functional integral is identically zero, we find that
\be \label{eq:pathid1}
    \Bigl\langle \bigl[ \lambda_0 (\sigma) \bigr]^n \, \bigl[ \overline{D} X_0 (\sigma) \bigr]^{\!n+1} \, \mathcal{O} \bigl[ \overline{\lambda}_0\,, x_0 \bigr](\sigma) \Bigr\rangle_{\!0} = 0\,.
\ee
Similarly, we have
\begin{align} \label{eq:pathid2}
    & \quad \Bigl\langle \bigl[ \overline{\lambda}_0 (\sigma) \bigr]^n \, \bigl[ {D} \overline{X}_0 (\sigma) \bigr]^{\!n+1} \, {\mathcal{O}} \bigl[ {\lambda}_0\,, x_0 \bigr](\sigma) \Bigr\rangle_{\!0} = 0\,.
\end{align}

Next, consider the correlation function
\begin{align}
    & \quad \Bigl\langle \bigl[ \lambda_0 (\sigma) \, \overline{D} X_0 (\sigma) \bigr]^{\!n} \, \mathcal{O} \bigl[ \overline{\lambda}_0\,, x_0 \bigr](\sigma) \Bigr\rangle_{\!0} \notag \\[2pt]
    & = \int \! \mathscr{D} \lambda_0 \, \mathscr{D} \overline{\lambda}_0 \, \mathscr{D} x_0 \sqrt{G [x_0]} \, \bigg\{ \frac{\delta}{\delta \lambda (\sigma')} \Bigl( N\bigl[\lambda_0\,, x_0\bigr] (\sigma\,, \sigma') \, \mathcal{O} [ \overline{\lambda}_0\,, x_0] (\sigma) \, e^{-S[x_0]} \Bigr) \notag \\[2pt]
    & \hspace{6cm} + n! \, (4\pi\alpha')^{n} \, \bigl[ \delta^{(2)} (\sigma - \sigma') \bigr]^n \, e^{-S[x_0]} \bigg\}_{\sigma'=\sigma} \,,      
\end{align}
where
\be
    N [\lambda_0\,, x_0] (\sigma, \sigma') = - n! \sum_{r=1}^n \frac{(4\pi\alpha')^r \bigl[ \delta^{(2)} (\sigma - \sigma') \bigr]^{r-1}}{(n+1-r)!} \lambda_0 (\sigma) \bigl[ \lambda_0(\sigma) \, \overline{D} X_0 (\sigma) \big]^{n-r}\,.
\ee
Dropping the total functional derivative term, we find that
\be
    \Bigl\langle \bigl[ \lambda_0 (\sigma) \, \overline{D} X_0 (\sigma) \bigr]^{\!n} \, \mathcal{O} \bigl[ \overline{\lambda}_0\,, x_0 \bigr](\sigma) \Bigr\rangle_{\!0} \! = n! \, (4\pi\alpha')^n \, \Bigl\langle \bigl[ \delta^{(2)} (\sigma - \sigma') \bigr]^n \Bigr\rangle_{\!0} \Big|_{\sigma'=\sigma} \,.
\ee
Similarly, 
\be
    \Bigl\langle \bigl[ \overline{\lambda}_0 (\sigma) \, D \overline{X}_0 (\sigma) \bigr]^{\!n} \, {\mathcal{O}} \bigl[ {\lambda}_0\,, x_0 \bigr](\sigma) \Bigr\rangle_{\!0} \! = n! \, (4\pi\alpha')^n \Bigl\langle \, \bigl[ \delta^{(2)} (\sigma - \sigma') \bigr]^n \Bigr\rangle_{\!0} \Big|_{\sigma'=\sigma}\,.
\ee
Therefore,
\begin{align} \label{eq:pathid3}
    & \quad \,\, \Bigl\langle \bigl[ \lambda_0 (\sigma) \, \overline{D} X_0 (\sigma) \bigr]^n \, \bigl[ \overline{\lambda}_0 (\sigma) \, {D} \overline{X}_0 (\sigma) \bigr]^m \, \CO [x_0] (\sigma) \Bigr\rangle_{\!0} \notag \\[2pt]
    & =  \Bigl\langle \bigl[ \lambda_0 (\sigma) \, \overline{D} X_0 (\sigma) \bigr]^m \, \bigl[ \overline{\lambda}_0 (\sigma) \, {D} \overline{X}_0 (\sigma) \bigr]^n \, \CO [x_0] (\sigma) \Bigr\rangle_{\!0}\,.
\end{align}

\subsection{One-loop beta-functions} \label{sec:oneloopbeta}

In this subsection, we return to the one-loop contribution to the effective action \eqref{eq:ShortenSeff} and take into account the path integral identities in \S\ref{sec:hantihcond} to derive the beta-functions in the nonrelativistic string theory sigma model. 

Our discussion so far has not included the counterterms. In fact, $\langle \cdots \rangle_0$ in \eqref{eq:Ox0} is defined with respect to the physical part of the functional couplings $E_{\mu\nu} [x_0]\,, A_{\mu\nu} [x_0]$ and $\tau_\mu{}^A [x_0]$ in \eqref{eq:S00}. In addition to the one-loop contribution from $S_\text{1-loop}$ in \eqref{eq:ShortenSeff}, one will also need to include the counterterm action
\be \label{eq:contertermaction}
    S_\text{c} \equiv \frac{1}{4\pi\alpha'} \int d^2 \sigma \, \Bigl[ \p x_0^\mu \, \overline{\p} x_0^\nu \bigl( \delta^E_{\mu\nu} + \delta^A_{\mu\nu} \bigr) + \lambda_0 \, \overline{\p} x^\mu_0 \, \delta^\tau_\mu{} + \overline{\lambda}_0 \, {\p} x^\mu_0 \, \delta^{\overline{\tau}}_\mu{} \Bigr]\,.
\ee
Then, the one-loop path integral defined with respect to the background fields $\lambda_0 \,, \overline{\lambda}_0$ and $x_0$ is
\begin{align} \label{eq:pathintegraleff}
    \CZ_\text{1-loop} \equiv \bigl\langle e^{-S_\text{1-loop}-S_\text{c}} \bigr\rangle^{}_{0} = \int \mathscr{D} \lambda_0 \, \mathscr{D} \overline{\lambda}_0 \, \mathscr{D} x_0^\mu \sqrt{G[x_0]} \, e^{-S_\text{eff}}\,,
\end{align}
where
\be
    S_\text{eff} \equiv S [\lambda_0, \, \overline{\lambda}_0, \, x_0] + S_\text{1-loop} + S_\text{c}\,.
\ee
The path integral identities \eqref{eq:pathid1}, \eqref{eq:pathid2} and \eqref{eq:pathid3} derived in \S\ref{sec:hantihcond} impose two conditions at the action level in $S_\text{1-loop} + S_\text{c}$\,:
\begin{enumerate}

\item

First, in $S_\text{1-loop} + S_\text{c}$\,,  one can set $\overline{D} X_0 = D \overline{X}_0 = 0$ in operators that do not contain any $\lambda_0$ or $\overline{\lambda}_0$\,, which we prove as follows. We only need to focus on terms in the expansion of the path integral $\CZ_\text{1-loop}$ that receive contributions from the $(\lambda\,, \overline{\lambda})$-independent terms in $S_\text{1-loop} + S_\text{c}$\,. Moreover, we only need to examine the part that contains $\overline{D}X_0$ or $D\overline{X}_0$ in these $(\lambda\,, \overline{\lambda})$-independent operators. These terms can be mixed up with contributions from the $\lambda_0 \, \overline{D} X_0$ and $\overline{\lambda}_0 \, D \overline{X}_0$ operators in the expansion of $\CZ_\text{1-loop}$\,, giving rise to terms of the following form:
\be \label{eq:lambdanDXm0}
    \bigl\langle \lambda_0^n \, \bigl( \overline{D} X_0 \bigr)^m \, \CO[\overline{\lambda}_0\,, x_0] \bigr\rangle_{\!0}\,,
        \qquad
    \bigl\langle \overline{\lambda}_0^n \, \bigl( {D} \overline{X}_0 \bigr)^{m} \, \CO[\lambda_0\,, x_0] \bigr\rangle_{\!0}\,,
        \qquad
    m > n\,,
\ee
which are identically zero by \eqref{eq:pathid1} and \eqref{eq:pathid2}. This effectively sets 
\be \label{eq:DXDX0}
    \overline{D} X_0 = D \overline{X}_0 = 0\,,
\ee 
in the terms that are independent of $\lambda_0$ or $\overline{\lambda}_0$ at the action level in $S_\text{1-loop} + S_\text{c}$\,.

\item

Second, one can set $\lambda_0 \, \overline{D} X_0 = \overline{\lambda}_0 \, D \overline{X}_0$ in $S_\text{1-loop} + S_\text{c}$\,. This is because the $\lambda_0 \, \overline{D} X_0$ and $\overline{\lambda}_0 \, D \overline{X}_0$ operators only contribute the nonvanishing terms
\be \label{eq:lnDXm=}
    \bigl\langle \lambda_0^n \, \bigl( \overline{D} X_0 \bigr)^m \, \CO[\overline{\lambda}_0\,, x_0] \bigr\rangle_{0}\,,
        \qquad
    \bigl\langle \overline{\lambda}_0^n \, \bigl( {D} \overline{X}_0 \bigr)^m \, \CO[\lambda_0\,, x_0] \bigr\rangle_{0}\,,
        \qquad
    m = n
\ee
to the expansion of $\CZ_\text{1-loop}$\,,
%
in which case the identity \eqref{eq:pathid3} applies. This effectively sets at the action level in $S_\text{1-loop} + S_\text{c}$
\be \label{eq:lambdaDX}
    \lambda_0 \, \overline{D} X_0 = \overline{\lambda}_0 \, D \overline{X}_0\,.
\ee

\end{enumerate}

We will therefore impose conditions \eqref{eq:DXDX0} and \eqref{eq:lambdaDX} at the level of effective action accordingly from now on. Consequently, the counterterm action \eqref{eq:contertermaction} becomes
\begin{align} \label{eq:Slambda0x0}
    S_\text{c} = \frac{1}{4\pi\alpha'} \int & d^2 \sigma \Bigl[ D x^{A'}_0 \, \overline{D} x^{B'}_0 \bigl( \delta^E_{A'B'} + \delta^A_{A'B'} \bigr)
    + \lambda_0 \, \overline{D} X_0 \, \delta^{\tau'} \notag \\
    & \qquad\,\, + D X_0 \, \overline{D X}_0 \, \delta^\Theta + D X_0 \, \overline{D} x^{A'}_0 \delta^{\Theta}_{A'} + \overline{DX}_0 \, D x^{A'}_0 \, \delta^{\overline{\Theta}}_{A'} \notag \\[2pt]
    & \qquad\,\, + \lambda_0 \, \bigl( \overline{D} x^{A'}_0 \delta^\tau_{A'} + \overline{DX}_0 \, \delta^\tau \bigr) + \overline{\lambda}_0 \, \bigl( D x^{A'}_0 \delta^{\overline{\tau}}_{A'} + DX_0 \, \delta^{\overline{\tau}} \, \bigr) \Bigr]\,,
\end{align}
where the counterterms labeled by the superscripts are associated with the couplings
\begin{subequations} \label{eq:tautautauTheta4}
\begin{align}
    \tau' & \equiv \tfrac{1}{2} \bigl( \tau_0 + \tau_1 + \overline{\tau}_0 - \overline{\tau}_1 \bigr)\,, 
        &
    \Theta_{AA'} & \equiv \tfrac{1}{2} \bigl( E_{AA'} + \epsilon_A{}^B A_{BA'} \bigr)\,, \\[2pt]
    \tau & \equiv \tfrac{1}{2} \bigl( \tau_0 - \tau_1 \bigr)\,,
        &
    \Theta_{A'} & \equiv \Theta_{0A'} + \Theta_{1A'}\,, \\[2pt]
    \overline{\tau} & \equiv \tfrac{1}{2} \bigl( \overline{\tau}_0 + \overline{\tau}_1 \bigr)\,,
        &
    \overline{\Theta}_{A'} & \equiv \Theta_{0A'} - \Theta_{1A'}\,, \\[2pt]
    & &
    \Theta & \equiv - \tfrac{1}{4} \bigl( E_A{}^A - \epsilon^{AB} A_{AB} \bigr)\,.
\end{align}
\end{subequations}

Furthermore, for $S_\text{1-loop}$ in \eqref{eq:ShortenSeff}, there is no dependence on $\lambda_0$ or $\overline{\lambda}_0$ and thus one just needs to impose the condition $D \overline{X}_0 = \overline{D} X_0 = 0$ as in \eqref{eq:DXDX0}. In addition, applying Identity \ref{id:identity11}, \ref{id:identity2}, \ref{id:RM} and \ref{id:looooong} in Appendix \ref{app:ids} to \eqref{eq:ShortenSeff}, we find
\begin{align} \label{eq:Seffoneloop}
    S_\text{1-loop} = - \frac{1}{2\alpha'} \int \! d^2 \sigma \Bigl\{ DX_0 \, \overline{DX}_0 \, \delta \Theta & + DX_0 \, \overline{D} x^{A'}_0 \delta \Theta_{A'} + \overline{DX}_0 \, Dx_0^{A'} \delta \overline{\Theta}_{A'} \notag \\
        & + Dx^{A'}_0 \overline{D} x^{B'}_0 \bigl( \delta E_{A'B'} + \delta A_{A'B'} \bigr) \Bigr\}\,,
\end{align}
where
\begin{subequations}
\begin{align}
    \delta \Theta & = \frac{\alpha'}{4} \bigl[ a \, \epsilon^{AB} \, \CR_{AB} (M) - \CR_{A'A}{}^{AA'} (G) - \tfrac{1}{2} \epsilon^{AB} \nabla^{A'} \CF_{A'AB} \notag \\
    & \hspace{6.45cm} + \tfrac{1}{4} \CF_{A'B'A} \CF^{A'B'A} \, \bigr] \, \Delta (0)\,, \\[2pt]
    \delta \Theta_{AA'} & = - \frac{\alpha'}{2} \bigl[ \eta_{AB} \CR_{A'B'}{}^{BB'} (G) + \tfrac{1}{2} \epsilon_A{}^{B} \, \nabla^{B'} \CF_{A'B'B} + \tfrac{1}{4} \CF_{B'C'\!A} \, \CF^{B'C'}{}_{\!\!A'} \bigr] \, \Delta (0)\,, \\[2pt]
    \delta E_{A'B'} & = - \frac{\alpha'}{2} \bigl[ \CR_{A'C'}{}^{B'C'} \! (J) + \CR_{B'C'}{}^{A'C'} \! (J) + \tfrac{1}{2} \CF_{C'D'A'} \CF^{C'D'}{}_{B'} \bigr] \, \Delta (0) \,, \\[2pt]
    \delta A_{A'B'} & = - \frac{\alpha'}{2} \nabla^{C'} \CF_{A'B'C'} \, \Delta (0)\,.
\end{align}
\end{subequations}
We recall that $\Delta(0) = \frac{1}{2\pi} \log \Lambda + \cdots$ in \eqref{eq:Delta(0)}.
Indeed, there are enough local counterterms in \eqref{eq:Slambda0x0} to absorb the loop divergences.  
Using the counterterms in \eqref{eq:Slambda0x0} to cancel the log divergences, we can read off the beta-functions as $\beta(\tau') = \beta(\tau) = \beta(\overline{\tau}) = \beta(\tau_{A'}) = \beta(\overline{\tau}_{A'}) = 0$ and
\begin{align} \label{eq:betafncwodilaton0}
    \beta^\Theta 
    = \frac{\delta \Theta}{2 \pi \Delta (0)}\,,
        \quad
    \beta^\Theta_{AA'} 
    = \frac{\delta \Theta_{AA'}}{2 \pi \Delta (0)}\,,
        \quad%
    \beta^E_{A'B'} = \frac{\delta E_{A'B'}}{2 \pi \Delta (0)}\,, 
        \quad%
    \beta^A_{A'B'} = \frac{\delta A_{A'B'}}{2 \pi \Delta (0)}\,. 
\end{align}
The beta-functions are defined without including any ``running" of the projectors. For example,
$\beta^E_{A'B'}$ actually represents $E^\mu{}^{}_{\!A'} \, E^\nu{}^{}_{\!B'} \, \beta (E^{}_{\mu\nu})$\,.
Recall that we set $m_\mu{}^A = 0$ in the curvature two-forms here, and all the $m_\mu{}^A$ dependence has been shuffled into the field strength $\CF_{\mu\nu\rho}$\,. Furthermore, supplemented with appropriate transformations of $\lambda$ and $\overline{\lambda}$\,, the action \eqref{eq:smaction} is invariant under the St\"{u}ckelberg transformations \cite{Bergshoeff:2018yvt, stringyNClimit}
\begin{subequations} \label{eq:Stueckelburg}
\begin{align} 
    H_{\mu\nu} & \rightarrow H'_{\mu\nu} - \bigl( C_\mu{}^A \, \tau_\nu{}^{\!B} + C_\nu{}^A \, \tau_\mu{}^{\!B} \bigr) \, \eta_{AB}\,, \\[2pt]
    B_{\mu\nu} & \rightarrow B'_{\mu\nu} + \bigl( C_\mu{}^A \, \tau_\nu{}^{\!B} - C_\nu{}^A \, \tau_\mu{}^{\!B} \bigr) \, \epsilon_{AB}\,.
\end{align}
\end{subequations}
These transformations leave $\Theta\,, \Theta_{AA'}\,, E_{A'B'} = H_{A'B'}$ and $A_{A'B'} = B_{A'B'}$ invariant. The beta-functions in \eqref{eq:betafncwodilaton0} are therefore invariant under the replacements in \eqref{eq:Stueckelburg} particularly for $C_\mu{}^A = - m_\mu{}^A$\,.\,\footnote{When $a=1$\,, the beta-functions in \eqref{eq:betafncwodilaton0} can be written in terms of the curvature tensor $R^\mu{}_{\nu\rho\sigma}$\,, defined with respect to the connection coefficients \cite{stringyNC}
\be \label{eq:affineconnectionwom}
    \Gamma^\rho{}_{\mu\nu} = \frac{1}{2} \, E^{\rho\sigma} \bigl( \p_{\mu} E_{\sigma\nu} + \p_{\nu} E_{\sigma\mu} - \p_\sigma E_{\mu\nu} \bigr) + \frac{1}{2} \, \tau^{\rho\sigma} \bigl( \p_{\mu} \tau_{\sigma\nu} + \p_{\nu} \tau_{\sigma\mu} - \p_\sigma \tau_{\mu\nu} \bigr)\,,
\ee
where $\tau_{\mu\nu} \! = \! \eta_{AB} \, \tau_\mu{}^A\, \tau_\nu{}^{B}\!, \,\, \tau^{\mu\nu} \! = \! \eta^{AB} \tau^\mu{}_{\!A} \, \tau^\nu{}_{\!B}$ and $E^{\mu\nu} \! = \! E^\mu{}_{\!A'} \, E^\nu{}_{\!A'}$. Note that $m_\mu{}^A$ is absent in \eqref{eq:affineconnectionwom}.
Setting $C_\mu{}^A = - m_\mu{}^A$ in the replacement rules in \eqref{eq:Stueckelburg} amounts to applying
\be
    \tau^{\mu\nu} \rightarrow N^{\mu\nu} = \tau^{\mu\nu} - \bigl( E^{\mu\rho} \, \tau^\nu{}_{\!A} + E^{\nu\rho} \, \tau^\mu{}_{\!A} \bigr) m_\rho{}^A
\ee
in addition to $E_{\mu\nu} \rightarrow H_{\mu\nu}$ in \eqref{eq:affineconnectionwom}. Here, $N^{\mu\nu}$ is introduced such that $N^{\mu\rho} \tau_{\rho\nu} + E^{\mu\rho} H_{\rho\nu} = \delta^\mu_\nu$ holds. The resulting $\Gamma^\rho{}_{\mu\nu}$ is the same as \eqref{eq:affineconnectioncoeff} \cite{stringyNC}.} 
This allows us to rewrite \eqref{eq:betafncwodilaton0} as
\begin{subequations} \label{eq:betafncwodilaton}
\begin{align}
    \beta^\Theta & = - \frac{\alpha'}{4} \ls \eta^{AB} \bigl( R_{AB} (a) - \tfrac{1}{4} \CH_{A'B'A} \CH_{A'B'B} \bigr) - \epsilon^{AB} \bigl( - \tfrac{1}{2} \nabla^{A'} \CH_{A'AB} \bigr) \rs, \label{eq:betafncwodilatonTheta} \\[2pt]
    \beta^\Theta_{AA'} & = \frac{\alpha'}{2} \ls \bigl( R_{AA'} (a) - \tfrac{1}{4} \CH_{B'C'A} \CH^{B'C'}{}_{\!A'} \bigr) - \tfrac{1}{2} \, \epsilon_A{}^B \nabla^{B'} \CH_{A'B'B} \rs, \\[2pt]
    \beta^H_{A'B'} & =  \alpha' \bigl( R_{A'B'} (a) - \tfrac{1}{4} \CH_{A'C'D'} \CH_{B'}{}^{C'D'} \bigr)\,, \\[2pt]
    \beta^B_{A'B'} & = - \frac{\alpha'}{2} \nabla^{C'} \CH_{A'B'C'}\,,
\end{align}
\end{subequations}
where
\begin{subequations} 
\begin{align}
    R_{\mu\nu} (a) & \equiv - a \, \epsilon^A{}_B \, \tau_\mu{}^B \, \CR_{A\nu} (M) + \tau_{\mu A} \, \CR_{A'\nu}{}^{\!AA'} (G) - E_\mu{}^{B'} \, \CR_{A'\nu}{}^{\!A'B'} (J)\,, \label{eq:Riccia} \\[2pt] 
    \CH_{\mu\nu\rho} & \equiv \p_\rho B_{\mu\nu} + \p_\mu B_{\nu\rho} + \p_\nu B_{\rho\mu}\,,
\end{align}
\end{subequations}
and the curvature two-forms are understood to be the ones derived in \cite{stringyNC,stringyNClimit}, without setting $m_\mu{}^A$ to zero anymore. Now, $m_\mu{}^A$ is only contained in the curvature two-forms and the covariant derivatives, while $\CH_{\mu\nu\rho}$ is just the field strength of the Kalb-Ramond field $B_{\mu\nu}$\,. Requiring that $R_{\mu\nu} (a)$ is invariant under the string-Galilean boost symmetry uniquely fixes $a = 1$\,. Furthermore, setting $a = 1$ in \eqref{eq:Riccia} gives $R_{\mu\nu} (1) \equiv R^\rho{}_{\mu\rho\nu}$\,, with $R^\rho{}_{\mu\sigma\nu}$ the curvature tensor defined with respect to the affine connection \eqref{eq:affineconnectioncoeff} \cite{stringyNC}.

\subsection{Nonrenormalization theorem} \label{sec:nonre}

In the one-loop contribution \eqref{eq:Seffoneloop} to the effective action, we found that no ($\lambda_0\,, \overline{\lambda}_0$)-dependent operator is generated and that there are sufficient local counterterms to absorb all the log divergences. Importantly, the marginal $\lambda_0 \overline{\lambda}_0$ operator is not generated quantum mechanically at one-loop. However, if there exist divergent higher-order loop diagrams that generate the $\lambda_0 \overline{\lambda}_0$ operator in the effective action, then one will need to turn on this operator before any quantum calculation to ensure that there are enough local counterterms to absorb the divergent quantum corrections. If this happens, the worldsheet fields $\lambda_0$ and $\overline{\lambda}_0$ will not be Lagrange multipliers anymore. Instead, the $\lambda_0 \overline{\lambda}_0$ operator would deform the theory towards relativistic string theory \cite{Gomis:2019zyu}. This can be shown most manifestly by considering nonrelativistic string theory in flat spacetime,
\be
    S = \frac{1}{4\pi\alpha'} \int d^2 \sigma \, \bigl( \, \p x^{A'}_0 \, \overline{\p} x^{A'}_0 + \lambda_0 \, \overline{\p} X_0 + \overline{\lambda}_0 \, \p \overline{X}_0 \, \bigr)\,,
\ee
but deformed by the $\lambda_0 \overline{\lambda}_0$ term,
\be \label{eq:Slambdalambda}
    S_{\lambda \overline{\lambda}} = \frac{1}{4\pi\alpha'} \int d^2 \sigma \, U \, \lambda_0 \, \overline{\lambda}_0\,,
\ee
with a constant $U$\,.\footnote{Note that the interactions from promoting $U$ to be $x_0$-dependent will be classically marginal.} Now, the action $S + S_{\lambda\overline{\lambda}}$ is no more string-Galilean boost invariant but acquires the global Lorentzian boost symmetry.
Then, integrating out the non-dynamical fields $\lambda_0$ and $\overline{\lambda}_0$ in the action $S + S_{\lambda \overline{\lambda}}$\,, we obtain the equivalent action (up to integration by parts)
\begin{align} \label{eq:Srel}
    S_\text{rel.} &     
    	= \frac{1}{4\pi\alpha'} \int d^2 \sigma \, \Bigl( \p x^{A'}_0 \, \overline{\p} x^{A'}_0 + U^{-1} \, \eta_{AB} \, \p x^A_0 \, \overline{\p} x^B_0 \Bigr)\,.
\end{align}
Defining $x^\mu_0 = \bigl( x^A_0 / \sqrt{U}\,, x^{A'}_0 \bigr)$\,,
\eqref{eq:Srel} becomes the standard Polyakov string action in conformal gauge. Turning on interactions in \eqref{eq:Srel} gives rise to the sigma model that describes strings propagating in a Riemannian geometry. Therefore, to decide whether the nonrelativistic string theory sigma model is self-consistent, we need to understand whether any divergent quantum correction to $\lambda_0 \overline{\lambda}_0$ is generated at higher loops.

In the following, we show that there is \emph{no} quantum correction at all loops to any marginal ($\lambda_0\,, \overline{\lambda}_0$)-dependent operators. Therefore, the two-dimensional sigma model that we have been considering so far is self-consistent quantum mechanically. In addition, we will show that the coupling $\tau_\mu{}^A$ does not run at all loops. 

We start with the higher order expansion of the sigma model action \eqref{eq:eqsmaction}. To focus on quantum corrections to ($\lambda_0\,, \overline{\lambda}_0$)-dependent terms, we only need to consider the ($\lambda_0\,, \overline{\lambda}_0$)-dependent interactions in the action and take the Taylor expansion of $\lambda \, \overline{\p} x^\mu \, \tau_\mu$ and $\overline{\lambda} \, {\p} x^\mu \, \overline{\tau}_\mu$ in \eqref{eq:eqsmaction}. 
Recall that, in \eqref{eq:lambdas}, we introduced dependence on the affine parameter $s$ in $\lambda (s)$ and $\overline{\lambda} (s)$\,. 
Taking into account the conditions in \eqref{eq:nablalambdamuconds} and \eqref{eq:lambdamuconds}, 
we find that
\be
    \nabla_s \Bigl( \lambda (s) \, \tau_\mu [y(s)] \Bigr) = 0\,. 
\ee
Addition, applying \eqref{eq:idforexpansions} repetitively, we obtain for $n \geq 2$\,,\,\footnote{When $n=2$\,, \eqref{eq:nablans} reduces to
$\nabla^2_s \Bigl( \lambda (s) \, \overline{\p} x^\mu (s) \, \tau_\mu [y(s)] \Bigr) \Big|_{s=0} = \lambda(0) \, \overline{\p} x_0^\mu \, \tau_\mu \, R^\mu{}_{\rho\sigma\nu} \, \ell^\rho \, \ell^\sigma$\,.}
\be \label{eq:nablans}
    \nabla^n_s \Bigl( \lambda (s) \, \overline{\p} x^\mu (s) \, \tau_\mu [y(s)] \Bigr) \Big|_{s=0} = \lambda(0) \, \tau_\mu \sum_{k=0}^{n-2} \mathcal{T}_k^{\,\mu_1 \cdots \mu_k \nu} \, \nabla_{\mu_1} \cdots \nabla_{\mu_k} R^\mu{}_{\rho\sigma\nu} \, \ell^\rho \, \ell^\sigma\,,
\ee
with $\CT_k = O (\ell^k)$\,. The explicit form of $\mathcal{T}_k^{\mu_1 \cdots \mu_k\nu}$ for general $k$ is not relevant for the following discussion. By \eqref{eq:curvaturetensor2form}, we have $\tau_\mu \, R^\mu{}_{\rho\sigma\nu} = \tau_\rho \, \CR_{\sigma\nu} (M)$\,. Moreover, $\tau_\mu \nabla_{\!\nu} \, \CO^\mu = \nabla_{\!\nu} \bigl( \tau_\mu \, \CO^\mu \bigr) + \Omega_\nu \, \tau_\mu \, \CO^\mu$\,. Applying these two identities to \eqref{eq:nablans}, we obtain 
\begin{align} \label{eq:allorderins}
    \nabla^n_s \Bigl( \lambda (s) \, \overline{\p} x^\mu (s) \, \tau_\mu [y(s)] \Bigr) \Big|_{s=0} 
        & = \lambda (0) \, \ell^\rho \, \ell^\sigma \, \sum_{k=0}^{n-1} \tilde{\mathcal{T}}_k^{\,\mu_1 \cdots \mu_k \nu} \, \nabla_{\mu_1} \cdots \nabla_{\mu_k} \bigl[ \tau_\rho \, R_{\sigma\nu} (M) \bigr] \notag \\[2pt]
        & = \lambda (0) \, \ell \, \ell^\sigma \, \sum_{k=0}^{n-1} \mathcal{T}_k^{\,\mu_1 \cdots \mu_k \nu} \, \nabla_{\mu_1} \cdots \nabla_{\mu_k} R_{\sigma\nu} (M)\,.
\end{align}
Using \eqref{eq:allorderins}, we obtain the Taylor expansion of $\lambda \, \overline{\p} x^\mu \, \tau_\mu$ as 
\begin{align} \label{eq:expandlambdaallloop}
    \lambda \, \overline{\p} x^\mu \, \tau_\mu [x] = \lambda_0 \, \overline{\p} x_0^\mu \, \tau_\mu & + \lambda_0 \, \overline{D} \ell + \sum_{n=2}^\infty \frac{1}{n!} \, \lambda_0 \, \ell \, \ell^\sigma \, \sum_{k=0}^{n-1} \mathcal{T}_k^{\,\mu_1 \cdots \mu_k \nu} \, \nabla_{\mu_1} \cdots \nabla_{\mu_k} R_{\sigma\nu} (M) \notag \\[2pt]
    + \rho \, \overline{\p} x_0^\mu \, \tau_\mu & + \rho \, \overline{D} \ell + \sum_{n=2}^\infty \frac{1}{n!} \, \rho \, \ell \, \ell^\sigma \, \sum_{k=0}^{n-1} \mathcal{T}_k^{\,\mu_1 \cdots \mu_k \nu} \, \nabla_{\mu_1} \cdots \nabla_{\mu_k} R_{\sigma\nu} (M)\,.   
\end{align}

\begin{figure}[t!] 
\centering
\begin{minipage}{4cm}
\begin{tikzpicture}
\draw [dotted] (2,2) circle (2 cm);
\draw (2,0) circle [radius=0.115];
\filldraw [white] (2,0) circle [radius=0.115];
\node at (2,0) {\scalebox{0.85}{$\times$}};
\draw (2,0) circle [radius=0.115];
\node  at (2.35,-.3) {$\lambda_0$};
\draw [thick] (2,0.115) to (1.5,1.25);
\node [scale = .8 ]at (1.7,.3) {$\ell$} ;
\draw [thick] (1.5,1.25) circle [radius= .035];
\filldraw [black] (1.5,1.25) circle [radius= .035];
\draw [thick] (1.5, 1.25) to (2.15, 2.25);
\draw [thick] (2.15, 2.25) circle [radius= .035];
\filldraw [black] (2.15, 2.25) circle [radius= .035];
\node [scale = .8] at (1.72, 1.15) {$\rho$};
\node [scale = .8] at (1.5, 1.55) {$\ell$};
\node [scale = .8] at (2.15, 2.00) {$\rho$};
\draw [thick] (1.72,3.98) -- (2.15, 2.25);
\filldraw [white] (2,3.2) circle [radius=0.4];
\draw [dotted, thick] (1.72,3.98) -- (2.15, 2.25);
\filldraw [white] (1.72,3.98) circle [radius=0.115];
\node at (1.72,3.98) {\scalebox{0.85}{$\times$}};
\draw (1.72,3.98) circle [radius=0.115];
\node at (2,4.3) {$\lambda_0$};
\node [scale = .8] at (1.95,3.7) {$\rho$};
\node [scale = .8] at (1.95, 2.4) {$\ell$};
\end{tikzpicture}
\end{minipage}
\caption{The subdiagram chain that starts and ends with two different insertions of vertices that contain a $\lambda_0$ field. The dotted circle denotes the Feynman diagram $\Upsilon$\,.}
\label{fig:chain}
\end{figure}
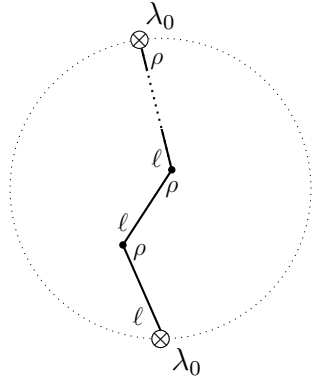

We attempt to construct a Feynman diagram $\Upsilon$ that contains an insertion of $\lambda_0$\,, which necessarily involve the vertices in \eqref{eq:expandlambdaallloop}. We make two observations: First, all vertices that involve $\lambda_0$ or $\rho$ necessarily involve an $\ell$ that contributes an internal leg in $\Upsilon$\,. Second, the only propagator that involves either $\rho$ or $\ell$ is $\langle 0 | \CT \rho (\sigma) \, \ell(\sigma') |0\rangle$ given in \eqref{eq:Grhoell}. Therefore, the Feynman diagram $\Upsilon$ necessarily contains a subdiagram that is a chain which starts and ends with two different insertions of vertices that both contain a $\lambda_0$ field. See Figure \ref{fig:chain}. 
The same arguments also apply to an $\overline{\lambda}_0$ insertion.  Hence, the $(\lambda_0\,, \overline{\lambda}_0)$-dependent operators that the $\Upsilon$ diagram corrects must contain a $\lambda_0 \, \lambda_0$ or $\overline{\lambda}_0 \, \overline{\lambda}_0$ factor,
which is forbidden by the rotational symmetry on the Euclidean worldsheet.\footnote{Note that $\lambda$ (and $\overline{\lambda}$) transforms nontrivially under the worldsheet rotation. In the infinitesimal case, we have $\delta \sigma^\alpha = \omega \, \epsilon^{\alpha}{}_{\!\beta} \, \sigma^\beta$\,, $\lambda = i \omega \lambda$ and $\overline{\lambda} = - i \omega \overline{\lambda}$\,.} Therefore, $\Upsilon$ has to vanish identically. 
This shows that there are no divergent or finite quantum corrections to any ($\lambda_0\,, \overline{\lambda}_0$)-dependent operators. When marginal operators are concerned, this implies that there is no quantum correction to the operators $\lambda_0 \overline{\lambda}_0\,, \lambda_0 \, \overline{D} X_0$ and $\overline{\lambda}_0 \, D \overline{X}_0$\,.

The above nonrenormalization theorem is a result of the hypersurface orthogonality condition \eqref{eq:curvatureconstraint}, required by the $Z_A$ gauge symmetry in \eqref{eq:deltaZ}. This point will be further elaborated on in Appendix \ref{app:torsion}.

\section{The Dilaton Contribution} \label{sec:dilaton}

In \S\ref{sec:quancal}, the beta-functions of the nonrelativistic string theory NLSM in a string Newton-Cartan geometry and Kalb-Ramond field background are computed. In this section, we consider the contributions from including the dilaton field in the sigma model action. 

On a curved worldsheet equipped with a metric $h_{\alpha\beta}$\,, $\alpha\,, \beta = 1\,, 2$\,, the action \eqref{eq:eqsmaction} becomes \cite{Bergshoeff:2018yvt}
\begin{align*}
    S & = S_\Phi + \frac{1}{4\pi\alpha'} \int d^2 \sigma \, \sqrt{h} \, \Bigl\{ \CD x^\mu \, \overline{\CD} x^\nu \bigl( E_{\mu\nu} [x] + A_{\mu\nu} [x] \bigr) + \lambda \, \overline{\CD} x^\mu \, \tau_{\mu} [x] + \overline{\lambda} \, \CD x^\mu \, \overline{\tau}_\mu [x] \Bigr\}\,,
\end{align*}
Here,
\be
    \CD \equiv \frac{i}{\sqrt{h}} \, \epsilon^{\alpha\beta} \, \overline{e}_\alpha \nabla_\beta\,,
        \qquad
    \overline{\CD} \equiv \frac{i}{\sqrt{h}} \, \epsilon^{\alpha\beta} {e}_\alpha \nabla_\beta\,.
\ee
We also included in the action the dilaton term,
\be
    S_\Phi = \frac{1}{4\pi} \int d^2 \sigma \, \sqrt{h} \, R [h] \, \Phi [x]\,,
\ee
with $R[h]$ the worldsheet Ricci scalar and $\Phi[x]$ the dilaton field.
The complication of working on a curved worldsheet can be avoided by focusing on the stress energy tensor. Even though the dilaton vanishes identically in the flat worldsheet limit, it contributes nontrivially the stress energy tensor,
\be
    T^{\Phi}_{\alpha\beta} = - \frac{4\pi}{\sqrt{h}} \frac{\delta}{\delta h^{\alpha\beta}} S_{\Phi} = \nabla_\alpha \p_\beta \Phi - h_{\alpha\beta} \nabla_\gamma \p^\gamma \Phi\,,
\ee
which in the flat limit $h_{\alpha\beta} \rightarrow \delta_{\alpha\beta}$ 
contributes
\be \label{eq:TPhi}
    T^\Phi [x] = - \p \overline{\p} \Phi [x]
\ee
to the trace anomaly. 
In the following, we compute the contributions from the trace anomaly \eqref{eq:TPhi} to the beta-functions in \eqref{eq:betafncwodilaton} on a flat worldsheet.

First, with respect to the background field $x_0$\,, the classical trace \eqref{eq:TPhi} is
\be \label{eq:TPhi[x0]2}
    T^\Phi [x_0] = - \p \overline{\p} x_0^\mu \, \p_\mu \Phi [x_0] - \p x^\mu_0 \, \overline{\p} x^\nu_0 \, \p_\mu \p_\nu \Phi[x_0]\,. 
\ee
Using the background field equation of motion\,\footnote{The equation of motion \eqref{eq:eomE} is derived by varying the linear $\ell^\mu$ terms in the action \eqref{eq:actioncovexpansion}. Here, it is more convenient to use the connection coefficients in the form of \eqref{eq:affineconnectionwom}.}
\begin{align} \label{eq:eomE}
    \bigl( \p \overline{\p} x^\rho_0 \, E_{\rho}{}^{A'} \bigr) E^\sigma{}_{\!A'} 
    & = \p x^\mu_0 \, \overline{\p} x^\nu_0 \bigl( \tau_\rho{}^{\!A} \, \tau^\sigma{}_{\!\!A} \, \Gamma^\rho{}_{\mu\nu} - \Gamma^\sigma{}_{\mu\nu} + \tfrac{1}{2} E^{\rho\sigma} \mathcal{F}_{\rho\mu\nu} \bigr)\,,
\end{align}
and the identity $\nabla \overline{\p} x_0^\mu \, \tau_\mu{}^A = 0$\,, we rewrite \eqref{eq:TPhi[x0]2} in a covariant form
\be \label{eq:TPhitree}
    T^\Phi [x_0] = - \p x_0^\mu \, \overline{\p} x^\nu_0 \bigl( \tfrac{1}{2} \, \mathcal{F}_{\!A'\mu\nu} \nabla_{\!A'} \Phi[x_0] + \nabla_{\!\mu} \nabla_{\!\nu} \Phi[x_0] \bigr)\,.
\ee
On the other hand, the trace of the stress energy tensor is related to the beta-functions $\beta_{\mu\nu}^H\,, \beta_{\mu\nu}^B\,, \beta_\mu^\tau$ and $\beta_\mu^{\overline{\tau}}$ as follows \cite{Gomis:2019zyu}:
\be \label{eq:tracestress}
    T^\alpha{}_\alpha [x_0] = - \frac{1}{2\alpha'} \Bigl[ \bigl( \beta^H_{\mu\nu} + \beta^B_{\mu\nu} \bigr) \p x^\mu \, \overline{\p} x^\nu + \beta^\tau_\mu \, \lambda \, \overline{\p} x^\mu + \beta^{\overline{\tau}}_\mu \, \overline{\lambda} \, \p x^\mu \Bigr]\,.
\ee
Comparing \eqref{eq:TPhitree} with \eqref{eq:tracestress}, we find that the beta-functions in \eqref{eq:betafncwodilaton} are extended with the dilaton contributions,
\begin{subequations} \label{eq:betafunctionswobetadilaton}
\begin{align}
    \beta^\Theta & = - \frac{1}{4} \, \alpha' \bigl( \eta^{AB} \, P_{AB} - \epsilon^{AB} \, Q_{AB} \bigr)\,, 
        &%
    \beta^H_{A'B'} & =  \alpha' \, P_{A'B'}\,, \\[2pt]
    \beta^\Theta_{AA'} & = \frac{1}{2} \, \alpha' \bigl( P_{AA'} + \epsilon_A{}^B \, Q_{BA'} \bigr)\,,
        &%
    \beta^B_{A'B'} & = \alpha' \, Q_{A'B'}\,,
\end{align}
\end{subequations}
where
\begin{subequations}
\begin{align}
   	{P}_{\mu\nu} & \equiv R_{\mu\nu} - \frac{1}{4} \, \mathcal{H}_{\mu A'B'} \mathcal{H}_\nu{}^{A'B'} + 2 \, \nabla_\mu \nabla_\nu \Phi\,, \\[2pt]
    	{Q}_{\mu\nu} & \equiv - \frac{1}{2} \nabla^{A'} \mathcal{H}_{A'\mu\nu} + \nabla^{A'} \Phi \, \mathcal{H}_{A'\mu\nu}\,.
\end{align}
\end{subequations}
Here, we applied \eqref{eq:Stueckelburg} to reshuffle the $m_\mu{}^A$ dependences in the final beta-functions. These resulting beta-functions match the ones derived in \cite{stringyNClimit,Gomis:2019zyu}.

So far, we have computed the contribution from the dilaton to the running of other couplings by working on a flat worldsheet. However, for the sigma model to be Weyl invariant, the trace of the stress energy tensor has to vanish on any curved worldsheet. The Weyl anomaly on a curved worldsheet can be computed in an elegant way by evaluating two-point functions of the stress energy tensor still on a flat worldsheet. A detailed explanation of this procedure for the string theory sigma model can be found in \cite{Callan:1989nz}, where the trace anomaly of the stress energy tensor is computed by evaluating $\langle T_{-+} T_{-+} \rangle$ on a flat worldsheet. Here, $T_{-+}$ is a component of the stress energy tensor in light-cone coordinates. Even to the lowest $\alpha'$-order in beta-functions, this calculation involves tree-level, one-loop and two-loop diagrams.

Application of the same method in \cite{Callan:1989nz} to the nonrelativistic string theory sigma model turns out to be almost identical to the relativistic case, except for replacing the relativistic propagator with $\langle 0 | \CT \, \ell^{A'} \! (\sigma) \, \ell^{B'} \! (\sigma') |0\rangle$ given by \eqref{eq:GA'B'} in the same nonvanishing Feynman diagrams. The resulting dilaton beta-function is \footnote{The beta-function $\beta^\Phi$ actually describes the running of $\Phi - \tfrac{1}{4} \ln G$\,, with $G$ given in \eqref{eq:det} \cite{Gomis:2019zyu}. }
\be \label{eq:betadilaton}
    \beta^{\Phi} = \frac{d- 26}{6} - \alpha' \bigl( \nabla_{A'} \nabla^{A'} \Phi - \nabla^{A'} \Phi \, \nabla_{A'} \Phi + \tfrac{1}{4} R_{A'}{}^{A'} - \tfrac{1}{48} \CH_{A'B'C'} \, \CH^{A'B'C'} \bigr)\,,
\ee
where $R_{A'}{}^{A'} = E^{\mu\nu} R_{\mu\nu}$\,. Here, $(d-26)/6$ comes from the central charge contribution from both the matter fields and the $bc$-ghosts in flat spacetime.
There are a few additional subtleties that are worth mentioning in this calculation in comparison with \cite{Callan:1989nz}. First, the stress energy tensor is different in nonrelativistic string theory because it receives contributions from the $(\lambda\,, \overline{\lambda})$-dependent operators in the action. The only change this brings is in the one-loop diagram that corrects $\langle T_{++} T_{++} \rangle$\,,\,\footnote{The contribution from the ($\lambda\,, \overline{\lambda}$)-dependent operators to $T_{++}$ on flat spacetime takes the form of $\lambda \, \p X$\,, expanding which with respect to $\rho$ and $\ell$ gives rise to a $\rho \, \p \ell$ vertex.} 
\be \label{eq:T++T++diagram}
\begin{tikzpicture}
\draw (0,0) to [out=90,in=90] (2,0);
\draw (0,0) to [out=-90,in=-90] (2,0);
\filldraw [white] (0,0) circle [radius=0.115];
\draw (0,0) circle [radius=0.115];
\node at (0,0) {\scalebox{0.85}{$\times$}};
\filldraw [white] (2,0) circle [radius=0.115];
\draw (2,0) circle [radius=0.115];
\node at (2,0) {\scalebox{0.85}{$\times$}};
\node at (-0.54,0) {\scalebox{0.9}{$\rho \, \p \ell$}};
\node at (2.54,0) {\scalebox{0.9}{$\rho \, \p \ell$}}; 
\end{tikzpicture} \notag \,
\ee
which is related to $\langle T_{-+} T_{-+} \rangle$ by imposing the conservation law of the stress energy tensor. This calculation gives rise to a central charge contribution from the matter fields. Second, when two-loop diagram contributions to the dilaton running are concerned, it is possible that there are diagrams  which contain vertices associated with the interaction operators, $\p \ell^\mu \, \overline{\p} \ell^\nu \, \ell^\rho \, \ell^\sigma \, \nabla_\rho \nabla_\sigma E_{\mu\nu}\,, \rho \, \ell \, \ell^A$ and $\overline{\rho} \, \overline{\ell} \, {\ell}{}^{A}$\,,\,\footnote{One can show that operators like $\rho \, \ell^A \ell^{A'}$ do not appear by using Identity \ref{id:RA'B'M} and \ref{id:RAA'M}.} that are not present in \cite{Callan:1989nz}\,. These interaction terms are obtained from the expansion of \eqref{eq:smaction} with respect to the quantum fields $\rho\,, \overline{\rho}$ and $\ell^\mu$\,. There is simply no two-loop diagram contributing to $\beta^\Phi$ that involves the vertices $\rho \, \ell \, \ell^A$ and $\overline{\rho} \, \overline{\ell} \, \ell^{A}$\,. 
However, the operator $\p \ell^\mu \, \overline{\p} \ell^\nu \, \ell^\rho \, \ell^\sigma \, \nabla_\rho \nabla_\sigma E_{\mu\nu}$ enters in the two-loop diagram
$$ 
\begin{tikzpicture}
\draw (0,0) to [out=90,in=90] (2,0);
\draw (0,0) to [out=-90,in=-90] (2,0);
\filldraw [white] (0,0) circle [radius=0.115];
\draw (0,0) circle [radius=0.115];
\node at (0,0) {\scalebox{0.85}{$\times$}};
\draw (0+2,0) to [out=90,in=90] (2+2,0);
\draw (0+2,0) to [out=-90,in=-90] (2+2,0);
\filldraw [white] (2+2,0) circle [radius=0.115];
\draw (2+2,0) circle [radius=0.115];
\node at (2+2,0) {\scalebox{0.85}{$\times$}};
\node at (-1,0) {\scalebox{0.9}{$\p \ell^{A'} \p \ell^{A'}$}};
\node at (2.35+2.75,0) {\scalebox{0.9}{$\p \ell^{A'} \p \ell^{A'}$}};
\node at (2,1) {\scalebox{0.9}{$\p \ell^\mu \, \overline{\p} \ell^\nu \, \ell^\rho \, \ell^\sigma \, \nabla_\rho \nabla_\sigma E_{\mu\nu}$}};
\filldraw [white] (2,0) circle [radius=0.115];
\draw (2,0) circle [radius=0.115];
\node at (2,0) {\scalebox{0.85}{$\times$}};
\end{tikzpicture}
$$
that corrects $\langle T_{++} T_{++} \rangle$\,. 
In this diagram, $\nabla_{\!\rho} \nabla_{\!\sigma} E_{\mu\nu}$ only contributes in the way with all indices contracted with $E^{\kappa\lambda}$\,, which turns out to be identically zero. 
This shows that the additional vertices do not contribute to $\beta^\Phi$ at the lowest $\alpha'$-order.

\section{Conclusion} \label{sec:conclusions}

In this paper, we developed a covariant background field method for string Newton-Cartan geometry and used it to compute beta functions of the NLSM that describes nonrelativistic string theory. The presence of the worldsheet fields $\lambda$ and $\overline{\lambda}$ that play the role of Lagrange multipliers in this NLSM adds a few interesting subtleties in this background field method calculation. First, in order to take into account necessary field redefinitions of $\lambda$ and $\overline{\lambda}$ that are required for the action to be covariant, we introduced a covariant expansion of these Lagrange multipliers with respect to the affine parameter. Second, by requiring that the string-Galilean boost symmetry hold in the beta-functions,
we identified the appropriate regularization of linearly divergent loop diagrams associated with the ($\lambda\,, \overline{\lambda}$)-dependent operators in the action. Finally, we proved a nonrenormalization theorem associated with the classical constraints imposed by integrating out the Lagrange multipliers $\lambda$ and $\overline{\lambda}$ in the path integral. 
After this analysis of the quantum effects from the presence of $\lambda$ and $\overline{\lambda}$\,, we applied a series of rather nontrivial identities in string Newton-Cartan geometry to derive the leading $\alpha'$-order beta-functions in \eqref{eq:betafunctionswobetadilaton} and \eqref{eq:betadilaton}, which are consistent with the results previously found in \cite{Gomis:2019zyu, stringyNClimit}. 

Having developed this machinery, we can more easily calculate higher loop corrections to the beta-functions, and hence higher derivative corrections to string Newton-Cartan gravity. It is also natural to extend the framework developed in this paper to include effects of D-branes and derive the nonrelativistic analogue of the Dirac-Born-Infeld effective field theory. 

\acknowledgments

We would like to thank Eric Bergshoeff, Jaume Gomis, Jihwan Oh, Jan Rosseel and Ceyda \c{S}im\c{s}ek for useful discussions. This research was supported in part by Perimeter Institute for Theoretical Physics. Research at Perimeter Institute is supported by the Government of Canada through the Department of Innovation, Science and Economic Development and by the Province of Ontario through the Ministry of Research, Innovation and Science.

\newpage

\appendix

\section{Identities in String Newton-Cartan Geometry} \label{app:ids}

In this appendix, we collect various identities in string Newton-Cartan geometry to facilitate the derivations in \S\ref{sec:oneloopbeta}.\,\footnote{In \cite{Aviles:2019xed}, string Newton-Cartan geometry in $2+1$ dimensions is studied using a Chern-Simons formulation, where some related identities are also discussed.} We first introduce a few more ingredients of string Newton-Cartan geometry, following \cite{stringyNC, stringyNClimit}.

We continue using the notation introduced in \S\ref{sectionwithFeynmanrules} around \eqref{eq:exsimnotation}, to exchange a curved index $\mu$ with flat indices $\{A, A'\}$ when it is contracted with an inverse Vielbein field when there is no derivatives acting on this inverse Vielbein. Also define
\be \label{eq:taumunu}
	\tau_{\mu\nu}{}^A \equiv \p_{[\mu} \tau_{\nu]}{}^A\,,
		\qquad%
	E_{\mu\nu}{}^{A'} \equiv \p_{[\mu} E_{\nu]}{}^{A'}\,.
\ee
As an example, we have $E_{AB'}{}^{A'} \equiv \tau^{\mu}{}_{\!A} \, E^\nu{}_{\!B'} \p_{[\mu} E_{\nu]}{}^{A'}$.
In terms of the this notation, the spin connections are \cite{stringyNC, stringyNClimit}
\begin{subequations}
\begin{align}
    \Omega_\mu & = \epsilon^{AB} \bigl( \tau_{\mu AB} - \tfrac{1}{2} \tau_\mu{}^C \tau_{ABC} \bigr)\,, \label{eq:scM} \\[4pt] 
    \Omega_\mu{}^{A A'} & = - E_{\mu}{}^{A A'} + E_{\mu B'}E^{A A' B'}\,, \label{eq:scG} \\[4pt]
    \Omega_\mu{}^{A'B'} & = - 2 E_\mu{}^{A'B'} + E_\mu{}^{C'} E^{A'B'}{}_{C'}\,.
\end{align}
\end{subequations}
It is also useful to define $\Omega_\mu{}^{A'A} = - \Omega_\mu{}^{AA'}$\,. We have set $m_\mu{}^A = 0$ in the spin connections, since $m_\mu{}^A$ is treated as a part of $A_{\mu\nu}$ in \eqref{eq:eqsmaction} after applying the field redefinitions of the Lagrange multipliers in \eqref{eq:lambdalambdatrans}. We have used the curvature two-forms $\CR_{\mu\nu} (M)\,, \CR_{\mu\nu}{}^{AA'} (G)$ and $\CR_{\mu\nu}{}^{A'B'} (J)$ to rewrite $R^\rho{}_{\sigma\mu\nu}$ as in \eqref{eq:curvaturetensor2form}. In terms of the spin connections $\Omega_\mu\,, \Omega_\mu{}^{AA'}$ and $\Omega_\mu{}^{A'B'}$\,, these curvature two-forms are \cite{stringyNC, stringyNClimit}
\begin{subequations}
\begin{align}	
    R_{\mu\nu} (M) & = 2 \p^{}_{[\mu} \Omega^{}_{\nu]}\,, \label{eq:RM}\\[4pt]
	R_{\mu\nu}{}^{A'B'} (J) & = 2 \bigl( \p^{}_{[\mu} \Omega_{\nu]}{}^{A'B'} + \Omega_{[\mu}{}^{A'C'} \Omega_{\nu]}{}^{B'}{}_{C'} \bigr)\,, \label{eq:RJ}\\[4pt]
	R_{\mu\nu}{}^{AA'} (G) & = 2 \bigl( \p^{}_{[\mu} \Omega^{}_{\nu]}{}^{AA'} + \epsilon^A{}_B \Omega_{[\mu}{}^{BA'} \Omega_{\nu]} + \Omega_{[\mu}{}^{AB'} \Omega_{\nu]}{}^{A'}{}_{B'} \bigr) \,. \label{eq:RG}
\end{align}
\end{subequations}
Finally, the curvature constraints \eqref{eq:curvatureconstraint} in our new notation become
\be \label{eq:A4}
    \tau_{\mu\nu}{}^A = \epsilon^A{}_B \, \Omega^{}_{[\mu} \tau^{}_{\nu]}{}^B\,,
        \qquad
    E_{\mu\nu}{}^{A'} = - \Omega^{}_{[\mu}{}^{AA'} \tau^{}_{\nu]A} + \Omega^{}_{[\mu}{}^{A'B'} E^{}_{\nu]}{}^{B'}\,.
\ee
From the first equation in \eqref{eq:A4}, we also obtain
\be \label{eq:fconst}
    \tau^{}_{A'AB} = - \tau^{}_{A'BA}\,,
        \qquad
    \tau^{}_{A'B'A} = 0\,. 
\ee

We now move on to derivations of a list of identities in string Newton-Cartan geometry. We will also derive a few related identities in the NLSM action, which requires imposing the constraints $\overline{D} X_0 = D \overline{X}_0 = 0$ in \eqref{eq:DXDX0}, already justified in \S\ref{sec:oneloopbeta}.

\vspace{1mm}

\begin{Def} \label{id:Omega[A'B']A}
    $\Omega_{[A'B']A} = 0$\,.
\end{Def}
\noindent \emph{Proof.} Plugging \eqref{eq:scG}, we find $\Omega_{[A'B']A} = \tfrac{1}{2} \bigl( E_{A'AB'} - E_{B'AA'} \bigr) - E_{A[B'A']} = 0$\,.

\vspace{1mm}

\begin{Def} \label{id:identity11}
    $\nabla_{[A'} E_{B']\mu} = 0$\,.
\end{Def}
\noindent \emph{Proof.} 
Using \eqref{eq:nablataunablaE}, we find
\be \label{eq:nablarhoEmunu}
    \nabla_\rho E_{\mu\nu} = - \eta_{AB} \, \Omega_\rho{}^{AA'} \bigl( E_\mu{}^{A'} \tau_\nu{}^B +  E_\nu{}^{A'} \tau_\mu{}^B \bigr)\,. 
\ee
Define $\Omega_\mu{}^{A'A} \equiv - \Omega_\mu{}^{AA'}$\,. Then,
$\nabla_{[A'} E_{B']\mu} = \tau_\mu{}^A \, \Omega_{[A'B']A}$\,.
Using Identity \ref{id:Omega[A'B']A} we find
$\nabla_{[A'} E_{B']\mu} = 0$\,.

\vspace{3mm}

\begin{Def} \label{id:identity2}
    $\int d^2 \sigma \, \overline{\nabla} \p x^\mu 
\, \nabla^{A'} \! E_{A'\mu} = 0$\,.
\end{Def}
\noindent \emph{Proof.} 
From \eqref{eq:nablarhoEmunu}, we have $\nabla^{A'} \! E_{A'\mu} = \tau_\mu{}^A \, \Omega^{A'}{}_{\!A'A}$\,. 
Define
\be \label{eq:OmegaOmegabar}
    \Omega_\mu{}^{A'} \equiv \frac{1}{2} \bigl( - \Omega_\mu{}^{A'0} - \Omega_\mu{}^{A'1} \bigr)\,,
        \qquad
    \overline{\Omega}_\mu{}^{A'} \equiv \frac{1}{2} \bigl( - \Omega_\mu{}^{A'0} + \Omega_\mu{}^{A'1} \bigr)\,.
\ee
Up to integration by parts, we have
\begin{align} \label{eq:nabladxnablaR}
    & \quad \int d^2 \sigma \, \overline{\nabla} \p x^\mu \, \nabla^{A'} \! E_{A'\mu} = \int d^2 \sigma \, \overline{\nabla} \p x^\mu \, \bigl( \tau_\mu \, \overline{\Omega}{}^{A'}{}_{\!A'} + \overline{\tau}_\mu \, \Omega{}^{A'}{}_{\!A'} \bigr) \notag \\
    & = - \int d^2 \sigma \, \overline{\p} x^\mu_0 \, \p x^\nu_0 \, \bigl( \nabla_\nu  \tau_\mu \, \overline{\Omega}{}^{A'}{}_{\!A'} + \tau_\mu \nabla_\nu \overline{\Omega}{}^{A'}{}_{\!A'} + \nabla_\mu \overline{\tau}_\nu \, \Omega^{A'}{}_{\!A'} + \overline{\tau}_\nu \nabla_\mu \Omega^{A'}{}_{\!A'} \bigr) \notag \\
    & = - \int d^2 \sigma \Bigl[ \, \overline{D} X_0 \, \p x^\mu_0 \, \bigl( - \Omega_\mu \! + \! \nabla_\mu \bigr) \, \overline{\Omega}{}^{A'}{}_{A'} + \overline{\p} x^\mu_0 \, D \overline{X}_0 \bigl( \Omega_\mu \! + \! \nabla_\mu \bigr) \, \Omega^{A'}{}_{\!A'} \Bigr] \notag \\[2pt]
    & = 0\,.
\end{align}

\vspace{1mm}

\begin{Def} \label{id:RA'B'M}
    $\CR_{A'B'} (M) = 0$\,.
\end{Def}
\noindent \emph{Proof.} Using \eqref{eq:RM}, we find
\begin{align} \label{eq:RA'B'M}
    \CR_{A'B'} (M) & = 2 \p_{[A'}\Omega_{B']}
    = 2 E^{[\mu}\,_{A'}E^{\nu]}\,_{B'} \,\epsilon^{A}{}_{B} \left(\p_\mu \tau^\rho{}_{\!A} \, \p_{[\nu}\tau_{\rho]}{}^B + \tau^\rho{}_{\!A} \, \p _\mu \p_{[\nu} \tau_{\rho]}{}^B \right)\,.
\end{align}
Differentiating both sides of the invertibility condition $\tau^\mu{}_{\!A} \, \tau_\nu{}^A + E^\mu{}_{A'} E_\nu{}^{A'} = 0$\,, we obtain
\begin{subequations} \label{eq:draisedind}
\begin{align}
    \p_\mu \tau^\rho{}_{\!A} & = - \tau^\sigma{}_{\!A} \bigl( \tau^\rho{}_B \, \p_\mu \tau_\sigma{}^B + E^\rho{}_{\!A'} \, \p_\mu E_\sigma{}^{A'}\bigr)\,, \label{eq:loweranIndex} \\[4pt]
    \p_\mu E^\rho{}_{\!A'} & = - E^\sigma{}_{\!A'} \bigl( \tau^\rho{}_B \, \p_\mu \tau_\sigma{}^B + E^\rho{}_{\!B'} \, \p_\mu E_\sigma{}^{B'}\bigr)\,. 
\end{align}
\end{subequations}
Applying \eqref{eq:loweranIndex} to \eqref{eq:RA'B'M}, we find
\begin{align}
    \CR_{A'B'} (M)
    & = - 2 \, E^{[\mu}{}_{A'} E^{\nu]}{}_{B'}\, \tau^\rho{}_{\!A} \, \Omega_\nu \, \p_{[\mu}\tau_{\rho]}{}^A \notag \\[2pt]
    & = - \bigl( \epsilon^A{}_B \, \tau^\rho{}_A \, \tau_\rho{} ^B \bigr) \bigl( E^{[\mu}{}_{A'}E^{\nu]}{}_{B'} \, \Omega_\mu \, \Omega_\nu \bigr) \notag \\[2pt]
    & = 0\,.
\end{align}

\vspace{1mm}

\begin{Def} \label{id:RAA'M}
    $\CR_{AA'} (M) = 0$\,. 
\end{Def}
\noindent \emph{Proof.} Using \eqref{eq:RM}, we find
\begin{align}
    \CR_{AA'} (M) & = 2 \, \tau^\mu{}_A E^\nu{}_{A'} \p_{[\mu} \Omega_{\nu]} \notag \\[2pt]
       & = \epsilon^{BC} \Bigl( 2 \, \tau^{}_{\rho (A C)} \, \p^{}_{A'} \tau^\rho{}^{}_B - \p^{}_{A'} \tau^{}_{A BC} + \p^{}_A \tau^{}_{A' BC} + \tau^{}_{A'AD} \, \tau^{}_{BC}{}^D \notag\\
    & \hspace{6cm} - \tau_{\rho A' C} \, \p^{}_{A} \tau^{\rho}{}_B + \tfrac{1}{2} \p^{}_{A'} \tau^{}_{B C A} \Bigr)\,. 
\end{align}
We emphasize that we only use the simplified notation when there are no derivatives acting on Vielbein fields with raised curved index $\mu$\,. For example, $\p_{A'} \tau_{ABC}$ is understood to be $\p_{A'} \tau_{ABC} = E^\mu{}_{\!A'} \, \tau^\nu{}_{\!A} \, \tau^\rho{}_B \, \p_{\mu} \p_{[\nu} \tau_{\rho] C}$\,.  
Further note that
\be \label{eq:epsilondt}
    \epsilon^{BC} \p^{}_{\!A'} \tau^{}_{BCA} = - 2 \, \epsilon^{BC} \bigl( \p^{}_B \tau^{}_{A'AC} + 2 \, \tau^{}_{\mu AC} \, \p^{}_B E^\mu{}_{A'} + \tau^{}_{A'\mu C} \, \p^{}_B \tau^\mu{}^{}_A + \tau^{}_{A' \mu A} \, \p^{}_B \tau^\mu{}_C \bigr)\,.
\ee
Using \eqref{eq:epsilondt} together with \eqref{eq:fconst} and \eqref{eq:draisedind}, we find
\be \label{eq:RAA'sim}
    \CR_{AA'} (M) = - 2 \, \epsilon^{BC} \, \tau^{}_{A'CD} \, \bigl( 2 \, \tau^{D}{}_{\!(AB)} + \tau^{}_{AB}{}^{D} \bigr)\,.
\ee
By repetitively using the Schouten identity for an arbitrary tensor $\CT_{ABC}$ with
\be
    \CT_{[AB]C} = \CT_{[CB]A} + \CT_{[AC]B}\,,
\ee
we find that
\begin{align*} 
    \epsilon^{BC} \tau^{}_{A' C}{}^D \, \tau^{}_{A B D} 
    & = \epsilon^{BC} \eta^{DE} \, \bigl( \tau^{}_{A'DE} \, \tau^{}_{A BC} + \tau^{}_{A'CD} \, \tau^{}_{AEB} \bigr) \, , \notag \\[2pt]
    & = \epsilon^{BC} \eta^{DE} \bigl( \tau^{}_{A'DA} \, \tau^{}_{EBC} + \tau^{}_{A'DB} \, \tau^{}_{AEC} + \tau^{}_{A'CD} \, \tau^{}_{AEB} \bigr) \notag \\
    & = \epsilon^{BC} \eta^{DE} \bigl( \tau^{}_{A'DB} \, \tau^{}_{EAC} + \tau^{}_{A'DC} \, \tau^{}_{EBA} + \tau^{}_{A'DB} \, \tau^{}_{AEC} + \tau^{}_{A'CD} \, \tau^{}_{AEB} \bigr)\,.  
\end{align*}
Using the first equation in \eqref{eq:fconst}, we obtain
\be \label{eq:CoolIdentity}
    \epsilon^{BC} \tau^{}_{A' C}{}^D \, \tau^{}_{A B D} = - 2 \, \epsilon^{BC} \eta^{DE} \tau^{}_{A'CD} \, \tau^{}_{E(AB)}\,.
\ee
Plugging \eqref{eq:CoolIdentity} back into \eqref{eq:RAA'sim}, we derive $\CR_{AA'} (M) = 0$\,.

\vspace{1mm}

\begin{Def} \label{id:RM}
    $\p x^\mu_0 \, \overline{\p} x^\nu_0 \, \CR_{\mu\nu} (M) = \frac{1}{4} DX_0 \, \overline{DX}_0 \, \epsilon^{AB} \, \CR_{AB} (M)$\,.
\end{Def}
\noindent \emph{Proof.}
First, we rewrite the $\CR_{\mu\nu} (M)$ term as follows:
\begin{align}
    \p x^\mu_0 \, \overline{\p} x^\nu_0 \, \CR_{\mu\nu} (M)
    & = DX_0 \, \overline{DX}_0 \, \tau^\mu \, \overline{\tau}^\nu \, \CR_{\mu\nu} (M)
    + D X_0 \, \overline{D} x_0^{A'} \tau^\mu E^\nu{}_{A'} \, \CR_{\mu\nu} (M) \notag \\[3pt]
    & \quad + D x^{A'}_0 \, \overline{D} x^{B'}_0 \, \CR_{A'B'} (M)
    + \overline{DX}_0 \, D x_0^{A'} \, \overline{\tau}^\mu E^\nu{}_{A'} \, \CR_{\mu\nu} (M)\,,
\end{align}
Then, using Identity \ref{id:RA'B'M} and \ref{id:RAA'M}, 
we find that
\begin{align} \label{eq:dxdxdOmega}
    \p x^\mu_0 \, \overline{\p} x^\nu_0 \, \CR_{\mu\nu} (M) & = \frac{1}{4} DX_0 \, \overline{DX}_0 \, \epsilon^{AB} \, \CR_{AB} (M)\,. 
\end{align}

\vspace{1mm}

\begin{Def} \label{id:looooong}
The following identity relates $\nabla_{\!\mu} \nabla_{\!\nu} \, E_{\rho\sigma}$ to the curvature two-form associated with the string-Galilean boost generator $G_{AA'}$\,:
\begin{align}
    & \quad \p x_0^\mu \, \overline{\p} x_0^\nu \bigl( \nabla^{A'} \nabla_{\!A'} E_{\mu\nu} - \nabla_\mu \nabla^{A'} \! E_{A'\nu} - \nabla_\nu \nabla^{A'} \! E_{A' \mu} \bigr) \notag \\[4pt]
    & = D X_0 \, \overline{D X}_0 \, \CO + D X_0 \, \overline{D} x^{A'} \CO_{A'} + \overline{D X}_0 \, D x^{A'} \, \overline{\CO}_{A'}\,,
\end{align}
where $\CO_{A'} \equiv \CO_{0A'} + \CO_{1A'}, \, \overline{\CO}_{A'} \equiv \CO_{0A'} - \CO_{1A'}$, and
\begin{subequations}
\begin{align}
    \CO = \frac{1}{2} \, \CR_{A'A}{}^{\!AA'} (G)\,,
        \qquad%
    \CO_{AA'} =  \eta_{A B}\CR_{A'B'}{}^{B B'} (G)\,.
\end{align}
\end{subequations}    
\end{Def}
\noindent \emph{Proof.} First, imposing the constraints $\overline{D} X_0 = D \overline{X}_0 = 0$\,,
we find
\begin{align} \label{eq:dxdxDDE}
    & \quad \p x_0^\mu \, \overline{\p} x_0^\nu \bigl( \nabla^{A'} \nabla_{\!A'} E_{\mu\nu} - \nabla_\mu \nabla^{A'} \! E_{A'\nu} - \nabla_\nu \nabla^{A'} \! E_{A' \mu} \bigr) \notag \\[4pt]
    & = D X_0 \, \overline{D X}_0 \, \CO + D X_0 \, \overline{D} x^{A'} \CO_{A'} + \overline{D X}_0 \, D x^{A'} \, \overline{\CO}_{A'}\,,
\end{align}
where 
\begin{subequations}
\begin{align}
    \CO & = 2 \, E^{\rho}{}_{\!A'} \, \tau^{\sigma}{}_{\!A} \p_{[\sigma} \Omega_{\rho]}{}^{A' A} \notag \\[2pt]
        & \quad + \frac{1}{2} \bigl( E^{\rho}{}_{\!A'} \, \tau^{\sigma}{}_{\!A} \nabla_{\rho} \Omega_\sigma{}^{A' A} - \epsilon^{AB} \Omega_A \, \Omega_{A'A'B} - \Omega_{A'B'}{}^{A} \, \Omega_{AA'B'} \bigr)\,, \\[6pt]
    \CO_{A'} & =  E^{\rho\sigma} \, \nabla_{\!\rho} \, \overline{\Omega}_\sigma{}^{A'} - E^\rho{}_{B'} E^\sigma{}_{A'} \nabla_{\!\rho} \, \overline{\Omega}_\sigma{}^{B'} \notag \\[2pt]
        & \quad + \overline{\Omega}_{B'C'} \, \Omega_{B'C'A'} 
	- \Omega_{B'} \, \overline{\Omega}_{B'A'} + \Omega_{B'} \, \overline{\Omega}_{A'B'} + \overline{\Omega}_{A'C'} \, \Omega_{B'B'C'} \,, \\[4pt]
    \overline{\CO}^{A'} & = E^{\rho\sigma} \, \nabla_\rho \, {\Omega}_\sigma{}^{A'} - E^\rho{}_{B'} E^\sigma{}_{A'} \nabla_\rho \, {\Omega}_\sigma{}^{B'} \notag \\[2pt]
        & \quad + {\Omega}_{B'C'} \, \Omega_{B'C'A'}  
	+ \Omega_{B'} \, {\Omega}_{B'A'} - \Omega_{B'} \, {\Omega}_{A'B'} + {\Omega}_{A'C'} \, \Omega_{B'B'C'}\,.
\end{align}
\end{subequations}
Here, $\Omega^{A'}$ and $\overline{\Omega}^{A'}$ are defined in \eqref{eq:OmegaOmegabar}. Also note that 
$\Omega_{A'B'} = E^\mu{}_{\!A'} \, \Omega_\mu{}^{B'}$ and $\overline{\Omega}_{A'B'} = E^\mu{}_{\!A'} \, \overline{\Omega}_\mu{}^{B'}$\,.
Then, it follows that
\begin{align}
    \CO_{AA'} & = E^\rho{}_{B'} E^\sigma{}_{A'} \nabla_{\!\rho} \Omega_\sigma{}^{B'}{}_{\!\!A} - E^{\rho\sigma} \nabla_{\!\rho} \Omega_{\sigma A'A} \notag \\[2pt]
        & \quad + 2 \, \epsilon_{AB} \, \Omega_{B'} \, \Omega_{A'B'}{}^B + \Omega_{B'A'C'} \, \Omega_{B'C'A} - \Omega_{B'B'C'} \, \Omega_{A'C'A}\,.
\end{align}

Using the explicit expression of the connection coefficients in \eqref{eq:affineconnectioncoeff}, we find
\begin{align} \label{eq:AfterSubThreeTerms}
     E^\rho{}_{\!A'} \, \tau^\sigma{}_{\!A} \nabla_{\!\rho} \Omega_\sigma{}^{A' A} & = 2 \, \bigl( \tau^{}_{A'AB} \, E^{}_{ABA'} + 2 \, E_{A'AB'} E^A{}_{(A'B')} \bigr) + \Omega_{A'B'}{}^A \, \Omega_{AA'B'}\,.
\end{align}
Using \eqref{eq:curvatureconstraint} to rewrite $\tau_{\mu\nu}{}^A$ and $E_{\mu\nu}{}^{A'}$ in terms of spin connections in \eqref{eq:AfterSubThreeTerms}, we obtain
\begin{align} \label{eq:FirstTermofHardExpand}
     E^\rho{}_{A'} \, \tau^\sigma{}_{\!A} \nabla_{\!\rho} \, \Omega_\sigma \, ^{A' A} & = - \epsilon^{AB} \, \Omega_{A'} \, \Omega_{AA'B}  + \Omega_{A'B'}{}^{A} \, \Omega_{AA'B'} \notag\\[2pt]
     & \quad - \tfrac{1}{2} \bigl( \Omega_{A'B'A} \, \Omega_{A'B'}{}^{A} + \Omega_{B'C'A} \, \Omega_{B'C'}{}^A \bigr)\,. 
\end{align}
Following the similar but a more lengthy procedure, we also find
\begin{align} \label{eq:FirstTermofHardExpand2}
    E^{\rho}{}_{B'} \, E^{\sigma}{}_{\!A'} \nabla_{\!\rho} \, \Omega_{\sigma B'A} - E^{\rho \sigma} \nabla_{\!\rho} \Omega_{\sigma A' A}  & = \Omega_{B' C' A} \, \Omega_{B' C' A'} + \Omega_{B' B' C'} \, \Omega_{A' C' A}\,.
\end{align}
Plugging \eqref{eq:FirstTermofHardExpand} and \eqref{eq:FirstTermofHardExpand2} in \eqref{eq:dxdxDDE}, and using \eqref{eq:RG}, we obtain
\begin{subequations}
\begin{align}
    \CO = \frac{1}{2} \, \CR_{A'A}{}^{\!AA'} (G)\,,
        \qquad%
    \CO_{AA'} =  \eta_{AB} \CR_{A'B'}{}^{BB'} (G)\,.
\end{align}
\end{subequations}

\section{Elements on Torsional String Newton-Cartan Geometry} \label{app:torsion}

In \S\ref{sec:nonre}, we showed that no $(\lambda\,, \overline{\lambda})$-dependent operator receives any quantum correction. We also mentioned at the end of \S\ref{sec:nonre} that this is a direct consequence of the hypersurface orthogonality condition $D_{[\mu} \tau_{\nu]}{}^A = 0$\,. To illustrate this point more explicitly, in the following, we consider the case when $D_{[\mu} \tau_{\nu]}{}^A \neq 0$ and show how the $\lambda \overline{\lambda}$ term is generated quantum mechanically already at one loop (see also \cite{Gomis:2019zyu}).\,\footnote{In this case, the $Z_A$ gauge symmetry is anomalous.} 

We start with imposing the Vielbein postulates as in \eqref{eq:vielbeinpostulates}, 
\begin{align}
    D_\mu \tau_\nu{}^A - \widehat{\Gamma}^\rho{}_{\mu\nu} \, \tau_\rho{}^A = 0\,, 
        \quad%
    D_\mu E_\nu{}^{A'} - \widehat{\Gamma}^\rho{}_{\mu\nu} \, E_\rho{}^{A'} = 0\,,
\end{align}
where the action of $D_\mu$ is defined in \eqref{eq:covariantdertauE}.
We only impose the curvature constraint $D_{[\mu} E_{\nu]}{}^A = 0$\,. It then follows that 
\be \label{eq:hatGamma}
    \widehat{\Gamma}^\rho{}_{\mu\nu} = {\Gamma}^\rho{}_{\mu\nu} + T^\rho_{\mu \nu}\,,
\ee
where ${\Gamma}^\rho{}_{\mu\nu} = \Gamma^\rho{}_{\nu\mu}$ is the torsionless part of the affine connection and
\be
   T^\rho{}_{\mu \nu} \equiv \tau^\rho{}_A \, D_{[\mu} \tau_{\nu]}{}^A\,, 
\ee
is the torsion. 
The covariant derivative acts on vector field $V_\mu$ and one-form field $\omega^\mu$ as
\be
    \widehat{\nabla}_{\!\mu} V_{\nu} = \p_\mu V_\nu - \widehat{\Gamma}^\rho{}_{\!\mu\nu} \, V_\rho\,,
        \qquad
    \widehat{\nabla}_{\!\mu} \, \omega^\nu = \p_\mu \, \omega^\nu + \widehat{\Gamma}^\nu{}_{\!\mu\rho} \, V^\rho\,.
\ee
Since we are focusing on quantum corrections to operators containing $\lambda_0$ and $\overline{\lambda}_0$\,, we only need to consider the covariant expansion of
\be \label{eq:Slambda,olambda}
    S_{\lambda} = \frac{1}{4\pi\alpha'} \int d^2 \sigma \, \lambda \, \overline{\p} x^\mu \, \tau_\mu\,,
\ee
and similarly for $S_{\overline{\lambda}}$\,,
with respect to the affine parameter $s$\,.
Even though the geodesic equation \eqref{eq:geodesicy} is not modified, the conditions \eqref{eq:nablalambdamuconds} on $\lambda_\mu (s)$ and $\overline{\lambda}_\mu (s)$ need to be modified to incorporate the torsion term, such that $\widehat{\nabla}_{\!s} \, \lambda_\mu (s) = \widehat{\nabla}_{\!s} \, \overline{\lambda}_\mu (s) = 0$\,.
At second order of the quantum fields $\ell^\mu$ and $\rho$\,, \eqref{eq:Slambda,olambda} gives 
\begin{align} \label{eq:lambdaXtau}
    S_{\lambda}^{(2)} & = \frac{1}{4\pi\alpha'} \int d^2 \sigma \, \Bigl\{ \rho \, \bigl( \overline{\nabla} \ell^\mu \, \tau_\mu - 2 \, \ell^\nu \, \overline{\p} x^\mu_0 \, \tau_{\!\rho} \, T^{\rho}{}_{\mu\nu} \bigr) - \tfrac{1}{2} \ell^\rho \ell^\sigma \lambda_0 \, \overline \p x^\mu \, \tau_\nu \widehat{R}^{\nu}{}_{\rho \mu \sigma} \notag \\[2pt]
    & \hspace{2.8cm} - \lambda_0 \Bigl[ \widehat{\overline{\nabla}} \ell^\rho \ell^\sigma \tau_\nu \, T^\nu{}_{\rho \sigma } + \overline{\p} x^\mu \ell^\rho \ell^\sigma \tau_\nu \bigl( \widehat{\nabla}_{\!\rho} \, T^\nu{}_{\!\mu\sigma} + 2 \, T^\nu{}_{\!\kappa \rho} \, T^\kappa{}_{\!\sigma\mu} \bigr) \Bigr] \Bigr\}\,.
\end{align}
Similarly, $S_{\overline{\lambda}}$ gives
\begin{align} \label{eq:lambdaXtaubar}
    S_{\overline{\lambda}}^{(2)} & = \frac{1}{4\pi\alpha'} \int \! d^2 \sigma \, \Bigl\{ \rho \, \bigl( {\nabla} \ell^\mu \, \overline{\tau}_\mu - 2 \, \ell^\nu \, {\p} x^\mu_0 \, \overline{\tau}_{\!\rho} \, T^{\rho}{}_{\mu\nu} \bigr) - \tfrac{1}{2} \ell^\rho \ell^\sigma \overline{\lambda}_0 \, \p x^\mu \, \overline{\tau}_\nu \widehat{R}^{\nu}{}_{\!\rho \mu \sigma} \notag \\[2pt]
    & \hspace{2.8cm} - \overline{\lambda}_0 \Bigl[ \widehat{\nabla}  \ell^\rho \ell^\sigma \, \overline{\tau}_\nu \, T^\nu{}_{\rho \sigma } + {\p} x^\mu \ell^\rho \ell^\sigma \overline{\tau}_\nu \bigl( \widehat{\nabla}_{\!\rho} \, T^\nu{}_{\!\mu\sigma} + 2 \, T^\nu{}_{\!\kappa \rho} \, T^\kappa{}_{\!\sigma \mu} \bigr) \Bigr] \Bigr\}\,.\!
\end{align}
Here, the curvature tensor $\widehat{R}^\mu{}_{\nu\rho\sigma}$ is defined to be
\be
    \widehat{R}^\mu{}_{\rho\sigma\nu} = \p_\sigma \widehat{\Gamma}^\mu{}_{\nu\rho} - \p_\nu \widehat{\Gamma}^\mu{}_{\sigma\rho} + \widehat{\Gamma}^\mu{}_{\sigma\kappa} \widehat{\Gamma}^\kappa{}_{\nu\rho} - \widehat{\Gamma}^\mu{}_{\nu\kappa} \widehat{\Gamma}^\kappa{}_{\sigma\rho}\,. 
\ee
Using \eqref{eq:lambdaXtau} and \eqref{eq:lambdaXtaubar} (together with other ($\lambda\,, \overline{\lambda})$-independent terms) we can derive as in \S\ref{sectionwithFeynmanrules} a set of Feynman rules for the vertices. In this set of rules, the ones that are relevant to quantum corrections to the $\lambda_0 \, \overline{\lambda}_0$ operator are
\begin{subequations}
\begin{align}
\begin{minipage}{3.8cm}
\begin{tikzpicture}
    \draw (0,0) -- (1.1,0) (1.3,0) -- (2.4,0);
    \draw (1.2,0) circle (1mm);
    \node at (1.2,0) {\scalebox{0.75}{$\times$}};
    \node at (1.2,0.36) {$\mathcal{T}$};
    \node at (0,-0.35) {$\ell^{A'} \! (\sigma)$};
    \node at (2.4,-0.35) {$\ell^{B'} \! (\sigma')$};
\end{tikzpicture}
\end{minipage}
    & = - \frac{1}{4\pi\alpha'} \int \mathcal{T}_{A'B'} (\sigma'') \, \bigl( \overline{\p}' - \overline{\p} \bigr) \, d\mu (\sigma\,, \sigma'\,, \sigma'')\,, \\[2pt]
\begin{minipage}{3.8cm}
\begin{tikzpicture}
    \draw (0,0) -- (1.1,0) (1.3,0) -- (2.4,0);
    \draw (1.2,0) circle (1mm);
    \node at (1.2,0) {\scalebox{0.75}{$\times$}};
    \node at (1.2,0.4) {$\overline{\mathcal{T}}$};
    \node at (0,-0.35) {$\ell^{A'} (\sigma)$};
    \node at (2.4,-0.35) {$\ell^{B'} (\sigma')$};
\end{tikzpicture}
\end{minipage}
    & = - \frac{1}{4\pi\alpha'} \int \overline{\mathcal{T}}_{\!A'B'} (\sigma'') \, \bigl( {\p}' - {\p} \bigr) \, d\mu (\sigma\,, \sigma'\,, \sigma'')\,,
\end{align}
\end{subequations}
where we defined $\mathcal{T}_{A' B'} = - {\lambda}_0 \, {\tau}_\mu T^{\mu}{}_{\!A' B'}$ and $\overline{\mathcal{T}}_{\!A' B'} =  - \overline{\lambda}_0 \, \overline{\tau}_\mu T^{\mu}{}_{\!A' B'}$\,.
There is a unique diagram that contributes the $\lambda_0 \overline{\lambda}_0$ corrections at one loop, 
\be \label{eq:doublelightballTT}
\begin{minipage}{3.3cm}
\begin{tikzpicture}
\draw (0,0) to [out=90,in=90] (2,0);
\draw (0,0) to [out=-90,in=-90] (2,0);
\filldraw [white] (0,0) circle [radius=0.115];
\draw (0,0) circle [radius=0.115];
\node at (0,0) {\scalebox{0.85}{$\times$}};
\filldraw [white] (2,0) circle [radius=0.115];
\draw (2,0) circle [radius=0.115];
\node at (2,0) {\scalebox{0.85}{$\times$}};
\node at (-0.35,0) {$\CT$};
\node at (2.35,0) {$\bar{\CT}$};
\end{tikzpicture}
\end{minipage}
= \frac{1}{4\pi} \, \log \Lambda \int d^2 \sigma \, \lambda_0 \, \overline{\lambda}_0 \, \tau_\mu \, \overline{\tau}_\nu \, T^{\mu}{}_{A'B'} \, {T}^\nu{}_{A'B'} + \text{finite}\,.
\end{equation}
However, there is no corresponding counterterm in \eqref{eq:Slambda0x0} to absorb this log divergence. Moreover, the appearance of the $\lambda_0 \overline{\lambda}_0$ operator invalidates the procedure of imposing the conditions \eqref{eq:DXDX0} and \eqref{eq:lambdaDX} at the level of the effective action. 
As expected, \eqref{eq:doublelightballTT} vanishes identically when the torsion $T^\rho{}_{\mu\nu}$ is set to zero. 

Note that \eqref{eq:lambdaXtau} and \eqref{eq:lambdaXtaubar} also contain vertices that contribute log divergent quantum corrections to the $\lambda_0 \, \overline{\p} x^\mu_0$ (and $\overline{\lambda}_0 \, \p x^\mu_0$) operators. Schematically, from \eqref{eq:lambdaXtau}, we can see that these quantum corrections at least contain terms like
\be \label{eq:a1a2}
    \lambda_0 \, \overline{\p} x^\mu_0 \, \bigl( a^{}_1 \, \tau_\nu \widehat{\nabla}_{A'} T^\nu{}_{\!\mu A'} + a^{}_2 \, \tau_\nu \, T^\nu{}_{\!\rho A'} \, T^\rho{}_{\!\mu A'} + \cdots \bigr) \log \Lambda\,.
\ee
The $\lambda_0 \, \overline{\p} x^\mu_0$ operator also receives other quantum corrections when other operators in the action are taken into account, such as 
\be \label{eq:a3}
    a_3 \, \lambda_0 \, \overline{\p} x^\mu_0 \, \tau_\nu \, T^\nu{}_{\!A'B'} \, \CH^{A'B'}{}_{\!\mu} \log \Lambda\,.
\ee
All these contributions again vanish identically when $T^\rho{}_{\mu\nu} = 0$\,. 

One may also consider deforming the NLSM by including the operator $\lambda \overline{\lambda}$ with a coupling $U$ together with all necessary counterterms and re-derive the beta-functions. It would then be interesting to look for solutions with $U=0$ while other couplings are tuned in the way such that these new beta-funcations vanish. In this case, there will be an extra equation from the vanishing beta-function for the coupling $U$ (even though $U$ itself is set to zero). Judging from the contributions given in \eqref{eq:doublelightballTT}, \eqref{eq:a1a2} and \eqref{eq:a3}, one expects that $\tau_\mu{}^A$ also runs and that $T^\rho{}_{\mu\nu}$ is generically nonzero in solutions to the spacetime equations of motion with $U=0$\,.\,\footnote{One may also start with integrating out the auxiliary fields $\lambda$ and $\overline{\lambda}$ in the path integral when $U\neq0$, which results in an action that is in form the same as the relativistic string NLSM but with a special parametrization of the functional couplings. Then, one can read off the spacetime equations of motion from requiring the Weyl invariance in relativistic string theory and consider a singular $U\rightarrow0$ limit there.} This scenario might be relevant to \cite{Harmark:2019upf}, where classical NLSMs in a torsional string Newton-Cartan background is proposed.
It will  also be interesting to compare this result with \cite{Gallegos:2019icg}, where Weyl invariance of the sigma model in a torsional Newton-Cartan background is considered.

\bibliographystyle{JHEP}
\bibliography{bgnrst}

\end{document}